\documentclass[twocolumn]{aa}
\usepackage{graphicx}
\usepackage{txfonts}
\usepackage{ulem}
\begin{document}

   \title{Contribution of mode coupling and phase-mixing of Alfv\'en waves to coronal heating}
   \titlerunning{Contribution of phase-mixing to coronal heating}

%   \subtitle{}

   \author{P. Pagano\inst{1}
          \and
          I. De Moortel\inst{1}
          }

   \institute{School of Mathematics and Statistics, University of St Andrews, North Haugh, 
St Andrews, Fife, Scotland KY16 9SS, UK.
              \email{pp25@st-andrews.ac.uk}
           }

   \date{ }

% \abstract{}{}{}{}{} 
% 5 {} token are mandatory
 
  \abstract
   {Phase-mixing of Alfv\'en waves in the solar corona has been identified as one possible candidate to explain coronal heating.
   While this scenario is supported by observations of ubiquitous oscillations in the corona carrying sufficient wave energy
   and by theoretical models that have described the concentration of energy in small scale structures,
   it is still unclear whether this wave energy can actually be converted into thermal energy in order to maintain the million degree solar corona.}
   {The aim of this work is to assess how much energy can be converted into thermal energy by a phase-mixing process
   triggered by the propagation of Alfv\'enic waves in a cylindric coronal structure, such as a coronal loop,
   and
   to estimate the impact of this conversion on the coronal heating and thermal structure of the solar corona.}
   {To this end, we run 3D MHD simulations of a magnetised cylinder
   where the Alfv\'en speed varies through a boundary shell and
   a footpoint driver is set to trigger kink modes
   which mode couple to
   torsional Alfv\'en modes in the boundary shell.
   These Alfv\'en waves are expected to phase-mix and the system allows 
   us to study the subsequent thermal energy deposition. We run a reference simulation to explain the main process and then we vary
   simulation parameters, such as the size of the boundary shell, its structure and the persistence of the driver.}
   {When we take into consideration high values of magnetic resistivity and strong footpoint drivers,
   we find
   i) that phase-mixing leads to a temperature increase of the order of $10^5$ K or less, depending on the structure of the boundary shell,
   ii) that this energy is able to balance the radiative losses only in the localised region involved in the heating,
   iii) and how the boundary layer and the persistence of the driver influence the thermal structure of the system.}
   {Our conclusion is that due to the extreme physical parameters we adopted and the moderate impact on the heating
   of the system, it is unlikely that phase-mixing can contribute on a global scale
   to the heating of the solar corona.}

   \maketitle
%
%________________________________________________________________

\section{Introduction}

The coronal heating problem has been open for several decades already
and it is still unclear what the mechanism is behind
the million degree solar corona.
Of course, all this time has not passed in vain,
and different generations of instruments and models 
have brought further insights to the problem.
We recommend \citet{DeMoortelBrowning2015} and references therein
for a comprehensive review.

One of the candidates to explain coronal heating is 
phase-mixing of Alfv\'en waves \citep{HeyvaertsPriest1983},
which is one of the main models where Alfv\'en waves
are put forward to explain the thermal structure of the solar corona.
\citep[see][for a more extended review]{Arregui2015}.
In this specific model localised small scale gradients develop
when Alfv\'en waves propagate
at different speed and these are preferred locations
where magnetic and kinetic energy can be converted into heating.
Recently, this model has fallen under careful scrutiny from the
coronal physics community because of theoretical results and 
observations that have opened the possibility for 
phase-mixing to explain coronal heating \citep[e.g. ][]{Cargill2016}.

Oscillations in the solar corona have been observed for more 
than a decade and some studies have already suggested
that the damping of waves could be connected with coronal heating
\citep{Nakariakov1999}.
However, only more recently observations of ubiquitous Alfv\'en waves
have concluded that waves carry enough energy to account for the coronal heating
\citep{Tomczyk2007,Jess2009,McIntosh2011},
where wave disturbances are sufficiently intense
to power quiet Sun and coronal holes,
while active regions remain off the range by an order of magnitude
\citep[see also][for a more comprehensive review]{DeMoortelNakariakov2012}.
\citet{MortonMcLaughlin2013} focused on low amplitude oscillations
where they found that wave activity is low over an extended period of time
and these would not be able to match the energy requirements of active regions.
\citet{Threlfall2013} identified unambiguous wave propagations
simultaneously measuring velocity and displacement of observed coronal loops
that \citet{LopezAriste2015} have interpreted as combined propagation of
kink and sausage wave modes.

At the same time, a number of studies have shown that transverse waves
are damped in the solar corona, opening the way for 
the possibility that the energy previously observed in the form of waves could be
converted into coronal heating.
\citet{Morton2014} analysed the power spectra of transverse motions in the solar corona and chromosphere
and have discovered a frequency-dependent transmission profiles motivated by the damping of kink waves in the lower corona.
\citet{Hahn2012} came to similar conclusions observing the line width of coronal lines in coronal holes,
adding that such damping could account for a major portion of the energy required to heat these structures.
\citet{Pascoe2016} observed and analysed the damping of some coronal loop oscillations
making a detailed seismologic analysis to retrive the loop parameters,
while \citet{Goddard2016} examined a large set of kink oscillations
to carry out a statistical study on how these oscillations undergo damping in the solar corona.
Additionally, other studies have confronted observed heating properties
with models in order to constraint the models.
\citet{VanDoorsselaere2007} found that coronal loop heating profiles better match with
a resistive wave heating mechanism than a viscous one.
\citet{Okamoto2015} observed the oscillations of threads of a coronal prominence,
idenfied as standing Alfv\'en waves,
and their subsequent damping and 
\citet{Antolin2015} argued that the observed damping of the 
oscillations is enhanced by the development of Kelvin-Helmotz Instabilities
at the boundaries of the threads.

On the theoretical side the highly structured solar corona suggests 
the presence of numerous interfaces between the many
magnetic structures where plasma and magnetic field vary
and offer the ideal environment where Alfv\'en waves can propagate at different speeds.
At the same time, such coronal structures are anchored to the base of the corona 
where footpoints move horizontally due to photospheric and chromospheric motions,
thus transversally to the magnetic field and likely to generate
transverse waves to be phase-mixed.
To this extend, the MHD numerical experiments of
\citet{Pascoe2010}, \citet{Pascoe2011}, \citet{Pascoe2012}, and \citet{Pascoe2013}
have shown that phase-mixing can be triggered in the solar corona
when kink oscillations of coronal loops
lead to the propagation of Alfv\'enic waves along the inhomogeneous flux tube.
It has been robustly proven that this process
leads to the concentration of wave energy in the inhomogeneous boundary shell
and the formation of small scale structures.
This model has also described how the structures generated by the phase-mixing 
on the boundary shell become increasingly smaller.
This result has been validated in different and increasingly realistic configurations 
and concluded that this energy needs to be eventually dissipated.
Earlier on, ideal MHD simulations by \citet{PoedtsBoynton1996} had estimated the heating due to phase-mixing of
Alfv\'en waves propagating along a magnetized cylinder by assuming turbulent heating following the ideal evolution.
\citet{Soler2016} have run a similar analysis of propagating Alfv\'en waves
in prominences
approaching the problem from an analytical point of view
and, estimating also the heating that can derive from this process,
they found that wave heating can contribute a fraction of the
radiative losses and only in certain conditions.

These data and numerical experiments have opened up the possibility
that the damping of kink oscillations could contribute to the coronal heating 
and with this regard phase-mixing is a likely candidate
mechanim that could enhance the damping of waves and lead to the conversion into coronal heating.
The aim of this work is therefore to test this hypothesis and
to assess the possible contribution to coronal heating that can derive 
from the damping of kink waves via phase mixing.

To this end we continue the study of \citet{Pascoe2010}
where MHD simulations of a magnetized cylinder 
with the presence of a driver at one of the footpoints are 
used to simulate the propagation of a kink mode in a coronal loop
and the subsequent mode coupling with the ($m=1$) Alfv\'en mode of the loop and dissipation of these waves through phase-mixing.
In our simulations we also account for non-ideal terms
as magnetic resistivity and thermal condution in order to 
investigate how much energy is converted into heating
and how the plasma temperature changes following this process.
To do so we first run a reference MHD simulation where a single pulse driver is set in 
and we analyse the energy deposition in the boundary shell and 
we then run a series of simulations where we investigate
the role of the width and shape of the boundary shell,
and how the results change when a continuous driver acts, instead of a single pulse.

The paper is structured as follows:
in Sec.\ref{model} we describe our reference model,
in Sec.\ref{simulation} we analyse in full details our reference simulation,
in Sec.\ref{parameterspaceinvestigation} we vary some properties of the boundary shell and the driver,
and in Sec.\ref{conclusions} we discuss our results and we draw some conclusions.

\section{Model}
\label{model}

In order to investigate the deposition of thermal energy in the solar corona
during the mode coupling and phase mixing we devise a numerical experiment 
following the same pattern introduced by \citet{Pascoe2010}, \citet{Pascoe2011}, \citet{Pascoe2012}.

\subsection{Initial Condition}
\begin{figure}
\centering
\includegraphics[scale=0.5,clip,viewport=80 68 370 645]{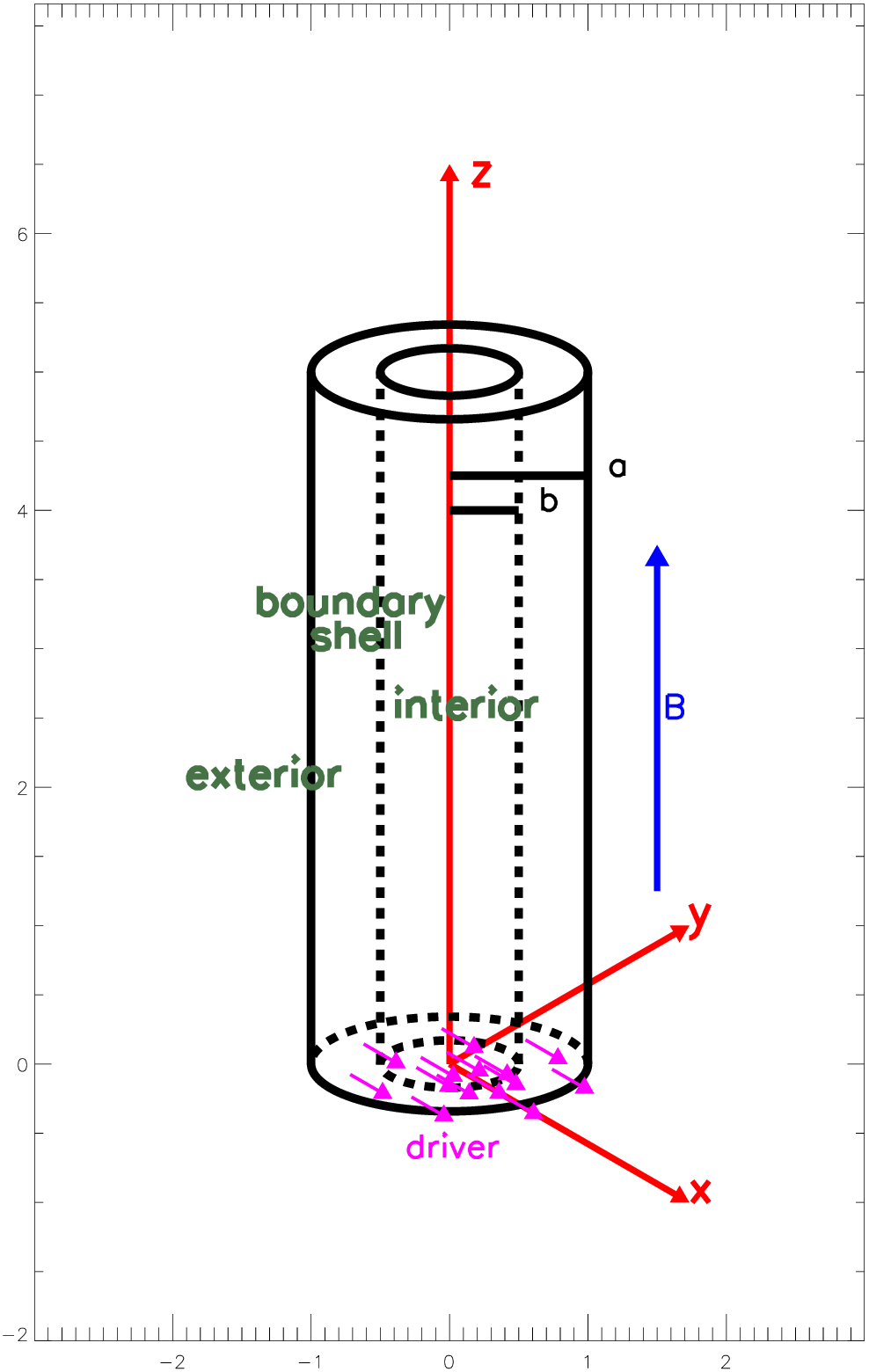}
\caption{Sketch to illustrate the geometry of our system and the cartesian axes.}
\label{sketch}
\end{figure}

We consider a cylindrical flux tube where
we define an interior region,
a boundary shell and an exterior region (Fig.\ref{sketch}).
The system is set in a cartesian reference frame 
with $z$ being the direction along the cylinder axis and 
$x$ and $y$ define the plane across the cylinder section.
The origin of the axes is placed at the centre of one footpoint of the cylinder.
The cylinder has radius $a$,
the interior region has radius $b$
and the boundary shell covers a fraction $l=\left(a-b\right)/a$ of the cylinder radius.
The boundary shell is the ring between radii $b$ and $a$,
and the exterior region is the rest of the domain beyond radius $a$.

The interior region is denser than the exterior 
and the density increases over the boundary shell
defined as a function of $\rho_e$, $\rho_i$, $a$, and $b$:
\begin{equation}
\label{densitylayer}
\displaystyle{\rho(\rho_e,\rho_i,a,b)=\rho_e+\left(\frac{\rho_i-\rho_e}{2}\right)\left[1-tanh\left(\frac{e}{a-b}\left[r-\frac{b+a}{2}\right]\right)\right]}
\end{equation}
where $r=\sqrt{x^2+y^2}$ is the radial distance from the centre of the cylinder,
$\rho_e$ is the density in the exterior region,
and $\rho_i$ is the density in the interior.
The temperature of the plasma, $T$, is assumed uniform at $T_0$,
and the thermal pressure, $p$, is set by the equation of state
\begin{equation}
\label{eos}
\displaystyle{p=\frac{\rho}{0.5 m_p} k_b T}
\end{equation}
where 
$m_p$ is the proton mass and $k_b$ is the Boltzmann constant.
The flux tube is initially in equilibrium and
in order to allow the propagation of kink waves and
provide magnetohydrostatic equilibrium we set
a non uniform magnetic field, $\vec{B}$, along the $z$-direction:
\begin{equation}
\label{bzlayer}
\displaystyle{B_z=\sqrt{B_e^2-2(p-p_e)}}
\end{equation}
that balances the varying thermal pressure,
where $B_e$ is the value of the field in the exterior region,
$p_e$ is the thermal pressure in the exterior region (derived from $\rho_e$ and $T_0$),
and $p$ is the local value of the thermal pressure.
In Tab.\ref{tableparameters} we list the values of the free parameters used to set up our specific experiment.
\begin{table}
\caption{Parameters}             % title of Table
\label{tableparameters}      % is used to refer this table in the text
\centering                          % used for centering table
\begin{tabular}{c c c}        % centered columns (4 columns)
\hline\hline                 % inserts double horizontal lines
Parameter & value & Units  \\    % table heading 
\hline                        % inserts single horizontal line
   $a$ & $1$ & $ Mm $  \\  
   $l$ & $0.5 $ & $ a $  \\  
   $\rho_i$ & $ 1.16\times10^{-16} $ & $g/cm^3$  \\  
   $\rho_e$ & $ 2.32\times10^{-16}$ & $g/cm^3$  \\  
   $T_0$ & $ 1.58$ & $MK$ \\
   $B_e$ & $ 6.18 $ & G \\

\hline                                   %inserts single line
\end{tabular}
\end{table}

\begin{figure}
\centering
\includegraphics[scale=0.5]{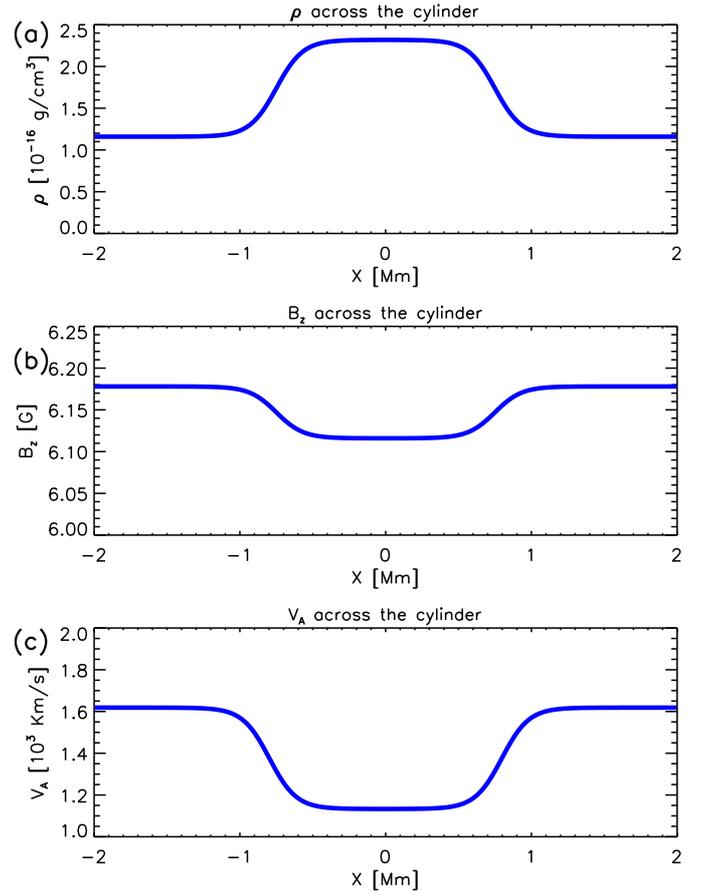}
\caption{Cuts of our initial condition for our MHD model
across the cylinder.
(a) density, 
(b) $z-$component of the magnetic field,
(c) Alfv\'en speed.}
\label{initialfig}
\end{figure}
Fig.\ref{initialfig} shows the value of $\rho$, $B_z$ and the 
corresponding Alfv\'en speed ($V_A$) across the cylinder.
In particular $V_A$ is higher in the exterior region and decreases 
in the boundary shell, where the maximum gradient is at $r\simeq0.8a$.
With the present parameters the plasma $\beta$ is uniformly $\beta=0.02$.

\subsection{The driver}
A driver acts on the system 
in order to initiate the propagation of Alfv\'enic waves within the system
(represented in pink in our sketch in Fig.\ref{sketch}).
The driver is prescribed by the displacement along the x axis of the flux tube 
below the lower $z$ boundary of the cylinder
and by the consequent velocity perturbation.
\begin{figure}
\centering
\includegraphics[scale=0.38]{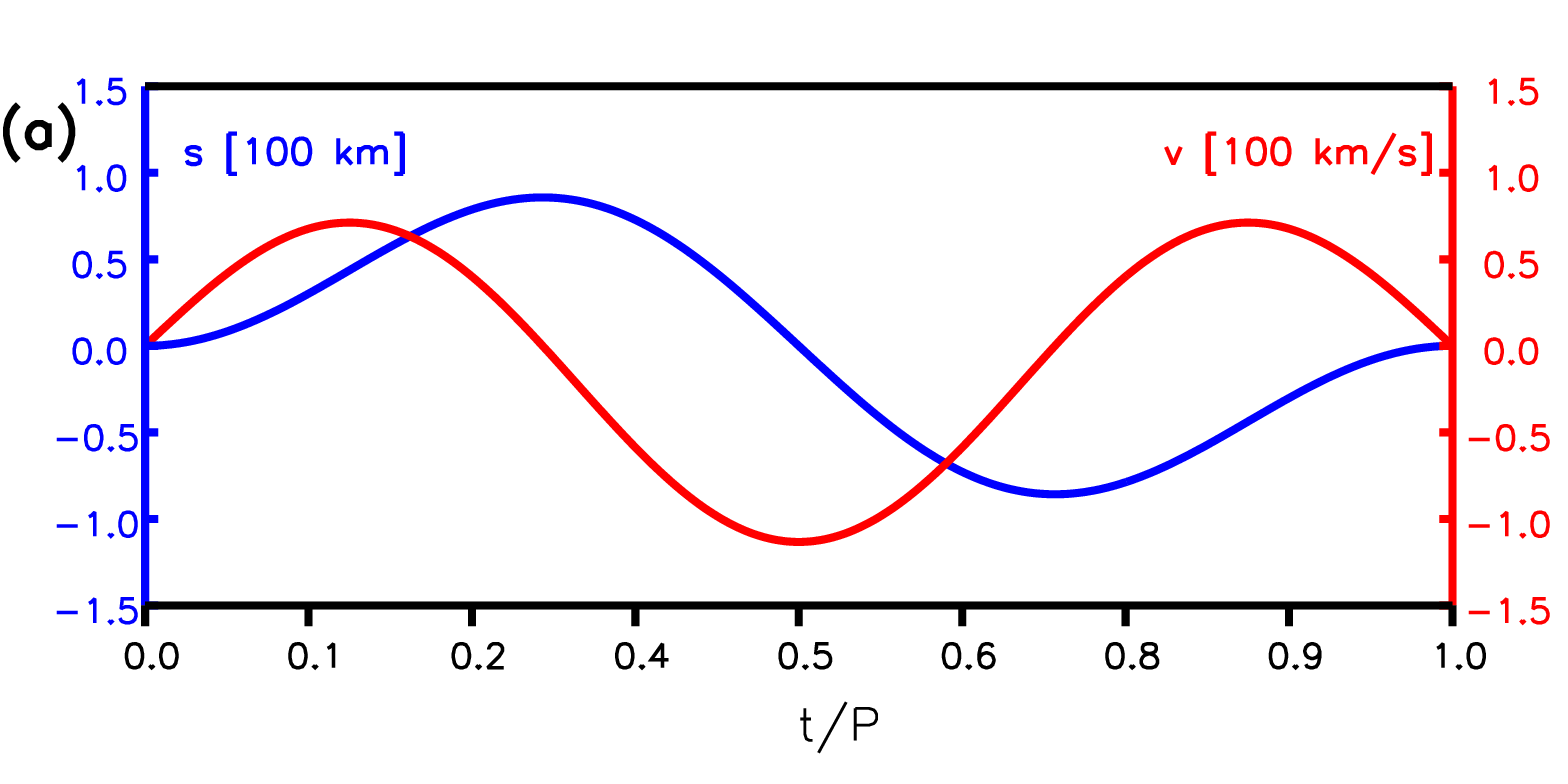} % printer
\includegraphics[scale=0.48]{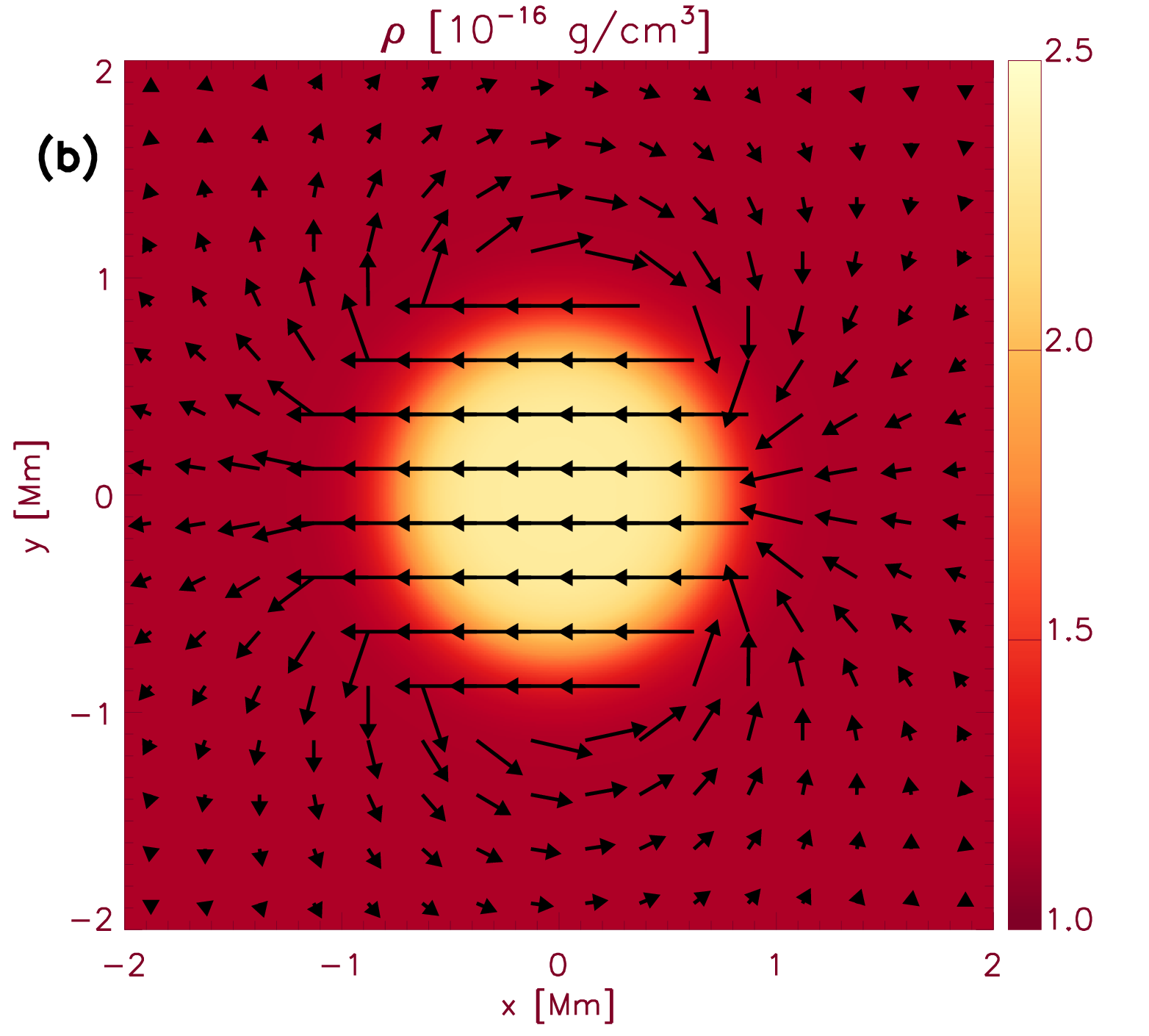} % printer
\caption{Settings for the driver.
(a) time evolution of the cylinder centre footpoint displacement as a function of time (blue line)
and the consequent velocity displacement (red line).
(b) map of the density of the driver at $t=0.5$ $P$ with overplotted velocity arrow.
The maximum velocity in the cylinder at this time is $V_0$ as from Tab.\ref{driverparameters}}
\label{figdriver}
\end{figure}
Specifically, Fig.\ref{figdriver}a (blue line) shows the position of
the centre of the flux tube as a function of time, whose expression is given by:
\begin{equation}
\label{sft}
\displaystyle{s(t)=\frac{V_0}{\omega}\sin{\omega t}\sin{\frac{\omega t}{2}}}
\end{equation}
where $V_0$ is the amplitude of the speed of the displacement, 
$\omega=2\pi/P$ is the angular frequency derived by the period of the oscillations $P$,
and $V_0/\omega$ is the following spatial displacement of the centre of the flux tube.
The drivers sets in at $t=0$ and stops at $t=P$.
Such a displacement of the flux tube leads to a uniform velocity perturbation 
within the interior region and the boundary shell that is simply described 
by the time derivative of Eq.\ref{sft} (Fig.\ref{figdriver}a, red line):
\begin{equation}
\label{vft}
\displaystyle{v_{x0}=\frac{ds(t)}{dt}=V_0\left(\cos{\omega t}\sin{\frac{\omega t}{2}}+0.5 \sin{\omega t}\cos{\frac{\omega t}{2}}\right)}.
\end{equation}
As in \citet{Pascoe2011}, this choice for the driver combines an oscillatory motion at the footpoint ($\sim \sin(\omega t)$) with an envelope of $\sin (\omega /2 t)$ to ensure a smooth (continuous) acceleration at $t=0$.
In the exterior region, we assume the system reacts to the flux tube motion as a
two dimensional dipole velocity configuration, so that the 
$x$ and $y$ components of the velocity in the exterior region can be written as function of time and space as:
\begin{equation}
\label{vxdrived}
\displaystyle{v_x=v_{x0} a^2 \frac{(x^2-y^2)}{(x^2+y^2)^2}}
\end{equation}
\begin{equation}
\label{vydrived}
\displaystyle{v_y=v_{x0} a^2 \frac{2xy}{(x^2+y^2)^2}}.
\end{equation}
This choice allows a smooth transition between the boundary shell and the exterior region
as the velocity amplitude is never discontinuous and mimics plasma flows around the magnetised cylinder when in motion.
Fig.\ref{figdriver}b shows the driver configuration at $t=P/2$ when the velocity
is maximum and the flux tube is at its rest position.
\begin{table}
\caption{Parameters}             % title of Table
\label{driverparameters}      % is used to refer this table in the text
\centering                          % used for centering table
\begin{tabular}{c c c}        % centered columns (4 columns)
\hline\hline                 % inserts double horizontal lines
Parameter & value & Units  \\    % table heading 
\hline                        % inserts single horizontal line
   $V_0$ & $ 113 $ & $ km/s $  \\  
   $P$ & $ 6.18 $ & $ s $  \\  
 
\hline                                   %inserts single line
\end{tabular}
\end{table}
Tab.\ref{driverparameters} summarises the values we have used to set up
the driver in our numerical experiment.

The value of $V_0$ we choose is one tenth of the Alfv\'en speed in
the interior and it is a relatively large value with respect 
to the usually considered horizontal foot point motions \citep{Threlfall2013}.
The period of the driver is designed to produce a visible effect on the present numerical experiment
and it is not meant to represent a significative frequency in the power spectrum of the corona.
At the same time, while the period is well below the peak of the 5 minutes,
oscillations of the order of seconds could still contribute to the power spectrum of
the velocity perturbation at the coronal footpoints.

\subsection{MHD simulation}

In order to study the evolution of the system due to the footpoint driver
we use the MPI-AMRVAC software~\citep{Porth2014} to solve the MHD equations,
where thermal conduction, magnetic diffusion and joule heating are treated as source terms:
\begin{equation}
\label{mass}
\displaystyle{\frac{\partial\rho}{\partial t}+\vec{\nabla}\cdot(\rho\vec{v})=0},
\end{equation}
\begin{equation}
\label{momentum}
\displaystyle{\frac{\partial\rho\vec{v}}{\partial t}+\vec{\nabla}\cdot(\rho\vec{v}\vec{v})
   +\nabla p-\frac{\vec{j}\times\vec{B}}{c}=0},
\end{equation}
\begin{equation}
\label{induction}
\displaystyle{\frac{\partial\vec{B}}{\partial t}-\vec{\nabla}\times(\vec{v}\times\vec{B})=\eta\frac{c^2}{4\pi}\nabla^2\vec{B}},
\end{equation}
\begin{equation}
\label{energy}
\displaystyle{\frac{\partial e}{\partial t}+\vec{\nabla}\cdot[(e+p)\vec{v}]=-\eta j^2-\nabla\cdot\vec{F_c}}, %-n^2\chi(T),
\end{equation}
where $t$ is time, $\vec{v}$ velocity,
$\eta$ the magnetic resistivity, $c$ the speed of light, $j=\frac{c}{4\pi}\nabla\times\vec{B}$ the current density,
$F_c$ the conductive flux \citep{Spitzer1962}.
The total energy density $e$ is given by
\begin{equation}
\label{enercouple}
\displaystyle{e=\frac{p}{\gamma-1}+\frac{1}{2}\rho\vec{v}^2+\frac{\vec{B}^2}{8\pi}}
\end{equation}
where $\gamma=5/3$ denotes the ratio of specific heats.

In the present numerical experiments we
adopt a value of $\eta$ which is set uniformly
as $\eta=10^9\eta_S$
where $\eta_S$ is the classical value \citep{Spitzer1962}.

The computational domain is composed of $512\times256\times512$ cells, distributed on a uniform grid.
The simulation domain extends from $x=-2$ $Mm$ to $x=2$ $Mm$,
from $y=-2$ $Mm$ to $y=0$ $Mm$ (where we model only half of a flux tube)
and from $z=0$ $Mm$ to $z=40$ $Mm$ in the direction of the initial magnetic field.
The boundary conditions are treated with a system of ghost cells
and we have periodic boundary conditions at both $x$ boundaries,
reflective boundary conditions at the $y$ boundary crossing the centre of the flux tube and
outflow boundary conditions at the other $y$ boundary.
The driver is set as a boundary condition at the lower $z$ boundary and outflow
boundary conditions are set at the upper $z$ boundary.

\section{Reference simulation}
\label{simulation}
\begin{figure}
\centering
\includegraphics[scale=0.28,clip,viewport=010 80 365 675]{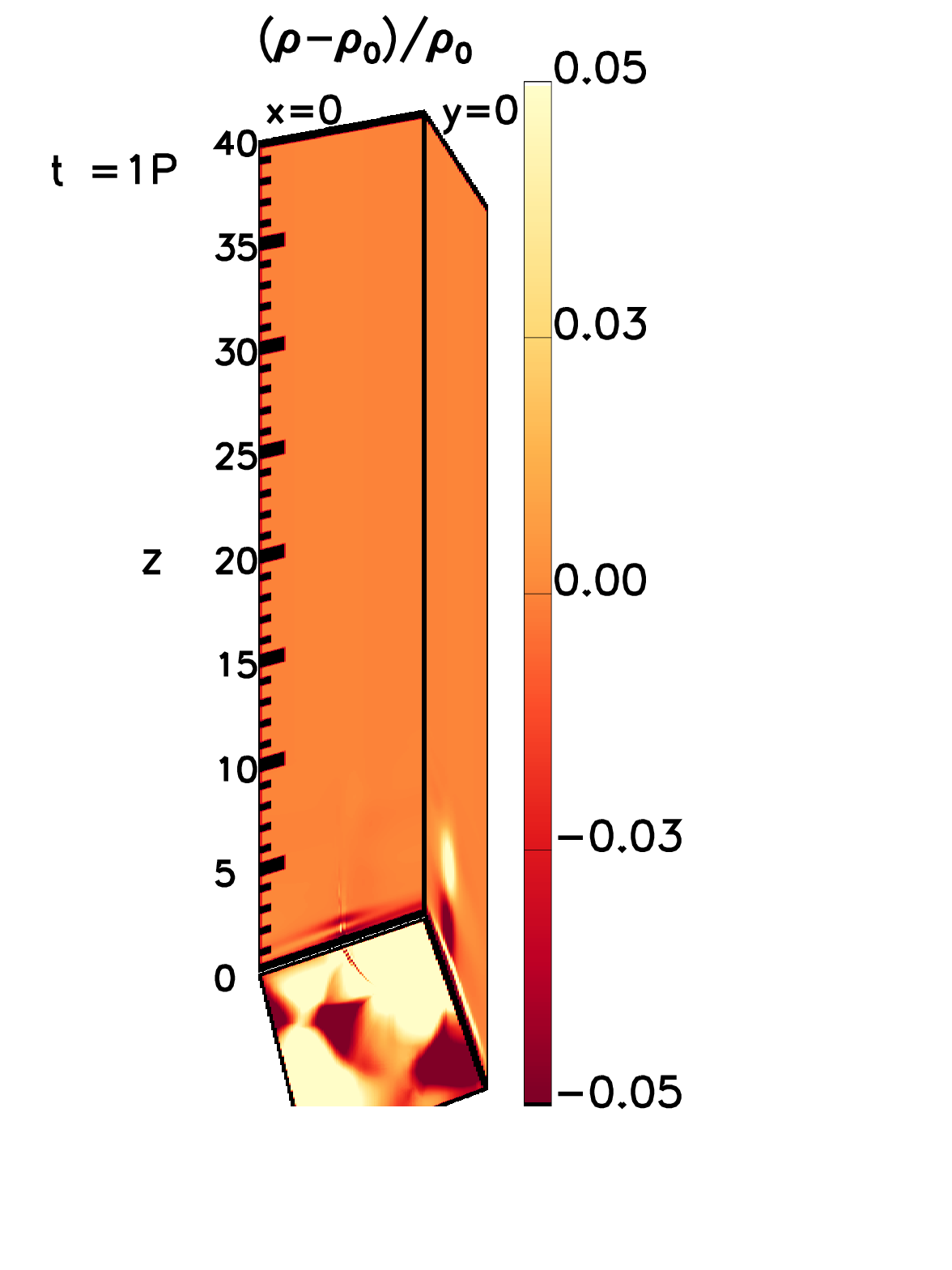} % printer
\includegraphics[scale=0.28,clip,viewport=110 80 365 675]{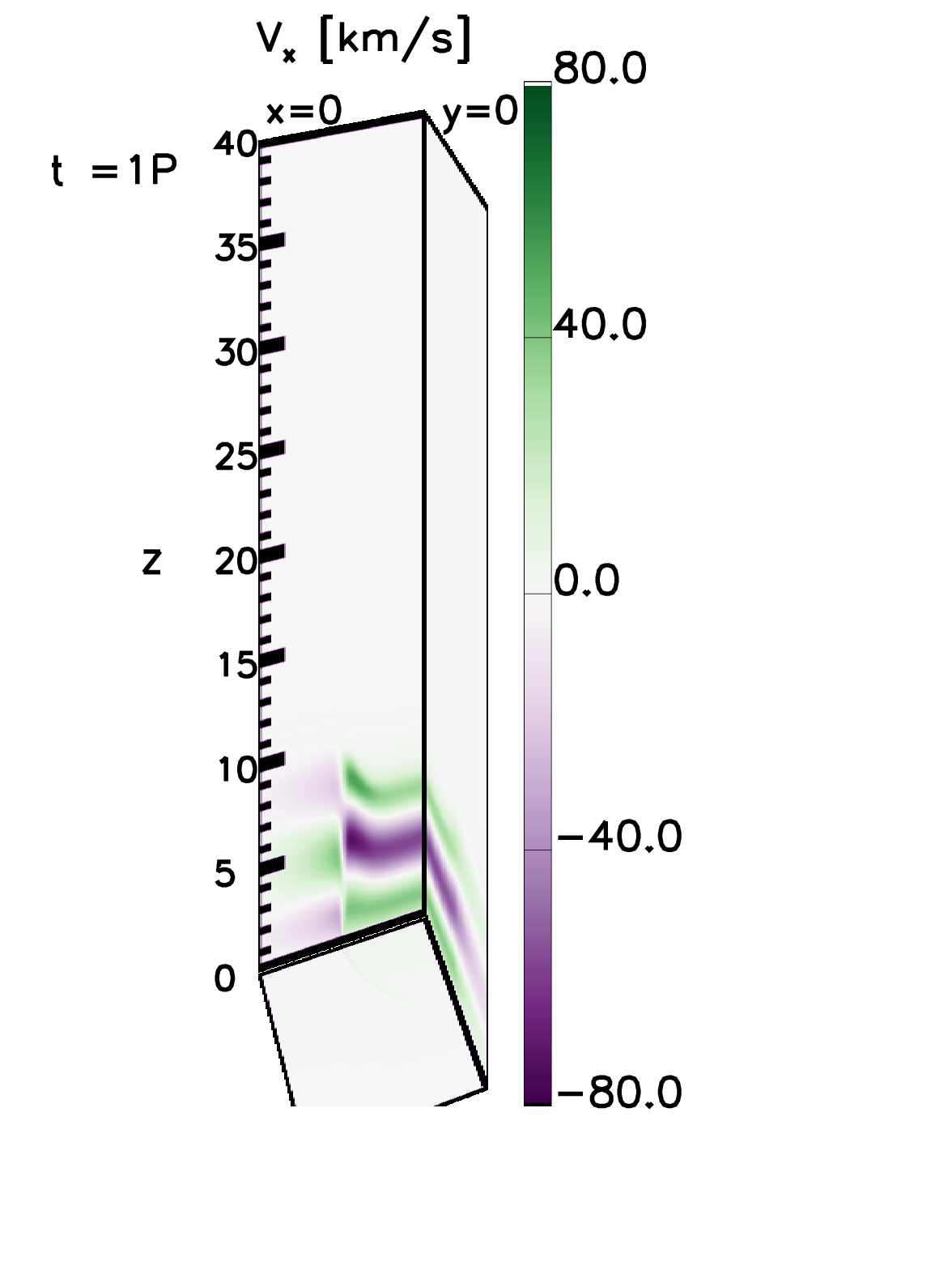} % printer
\includegraphics[scale=0.28,clip,viewport=110 80 365 675]{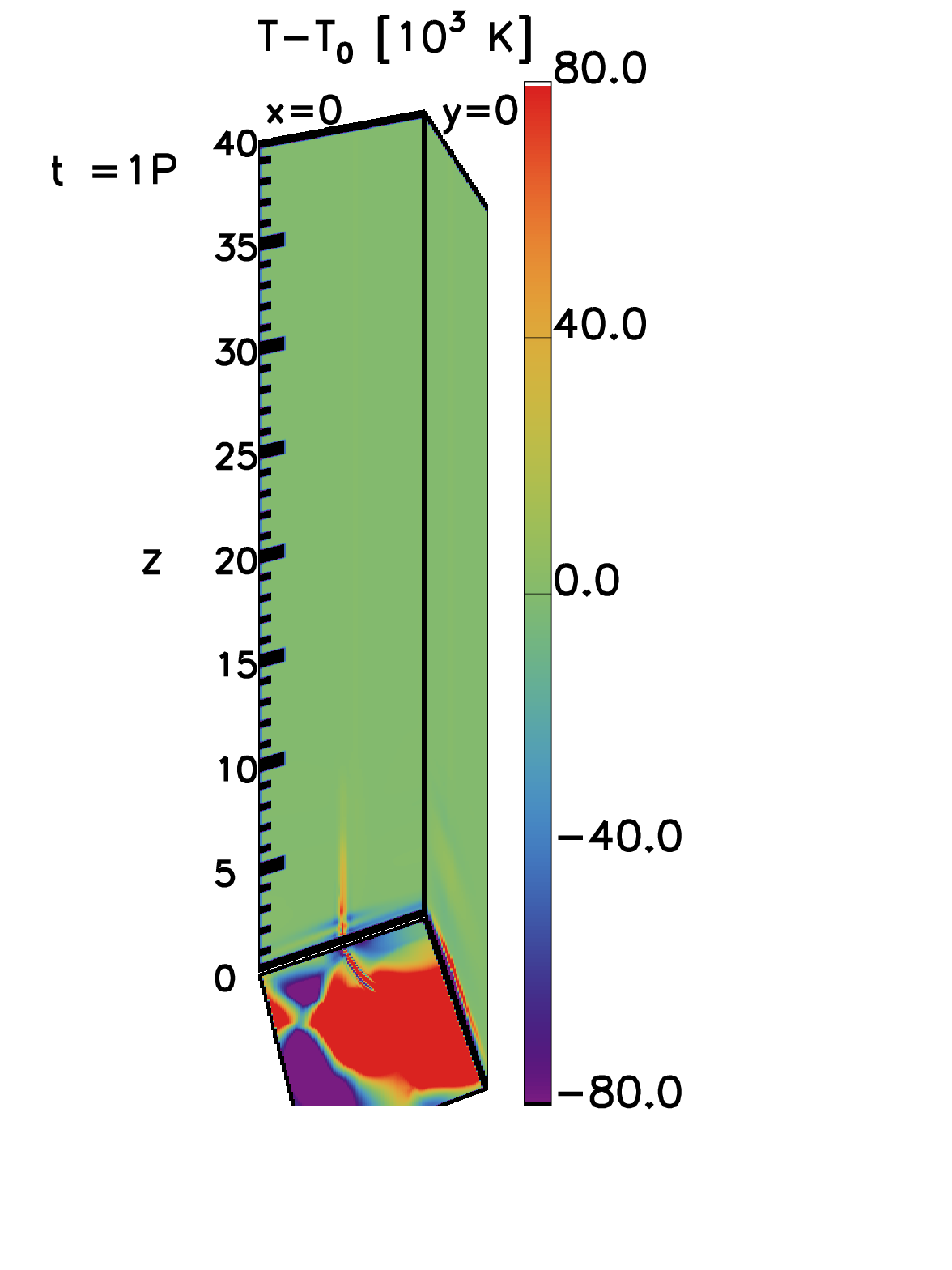} % printer

\includegraphics[scale=0.28,clip,viewport=010 80 365 675]{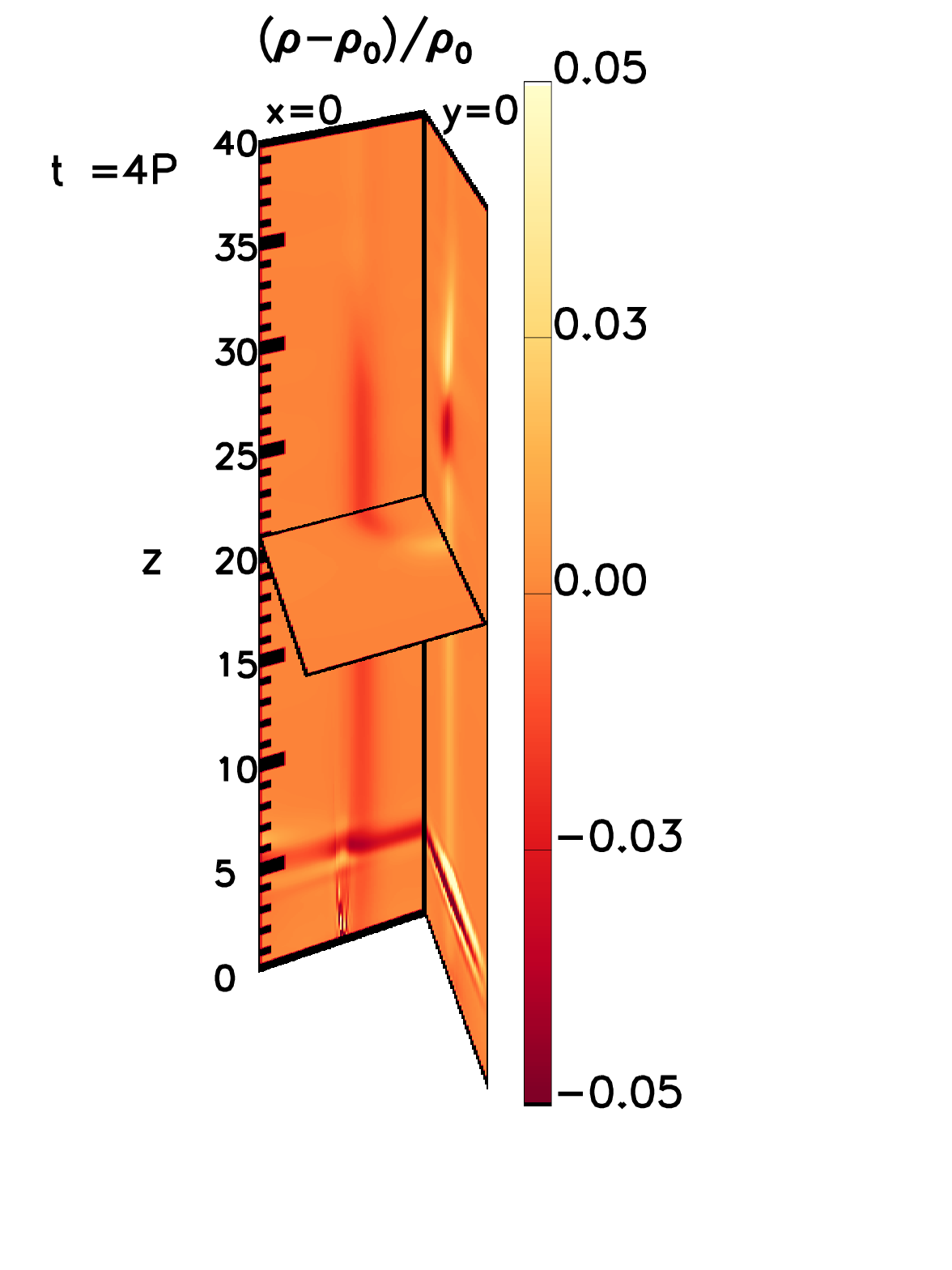} % printer
\includegraphics[scale=0.28,clip,viewport=110 80 365 675]{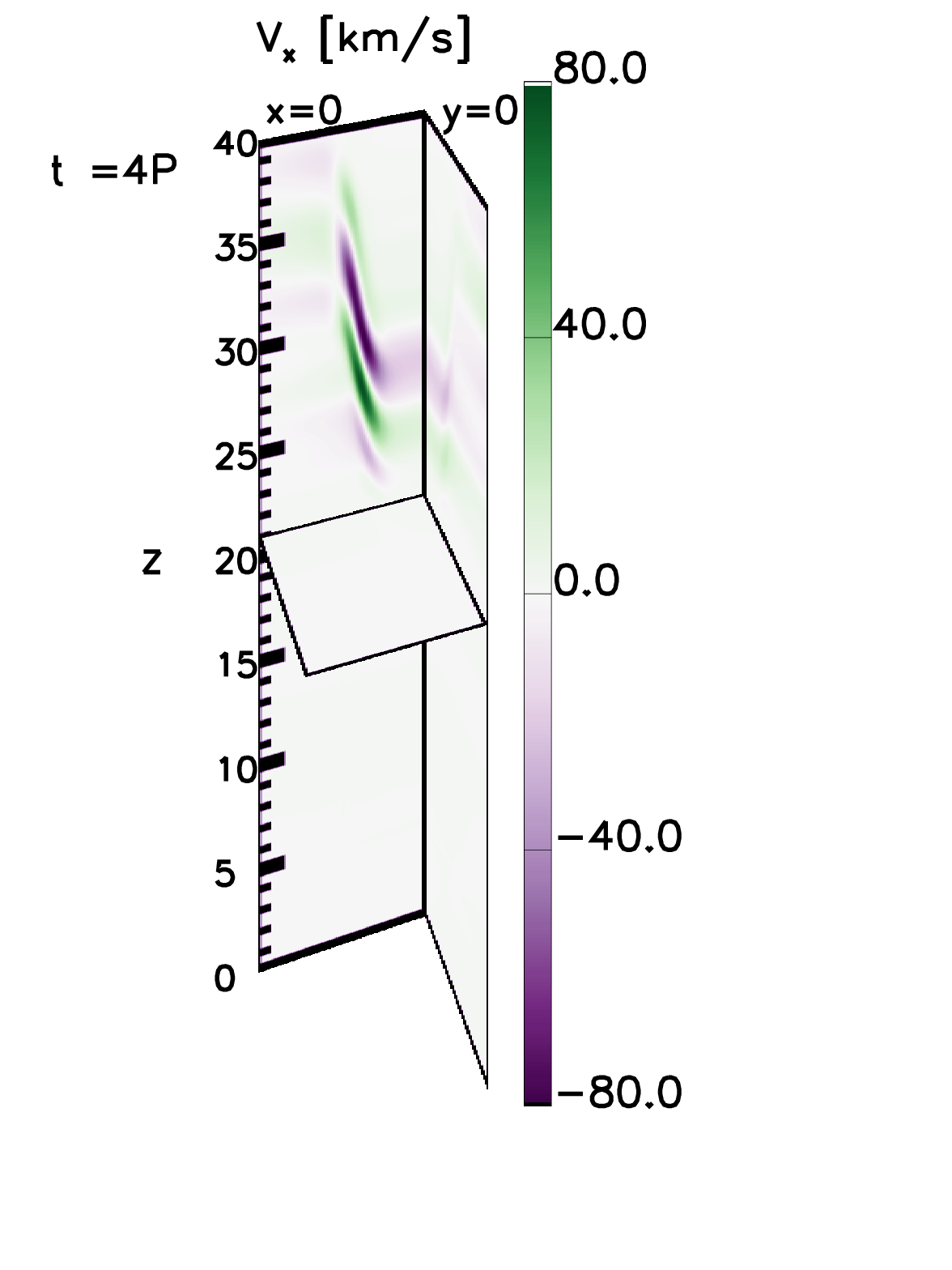} % printer
\includegraphics[scale=0.28,clip,viewport=110 80 365 675]{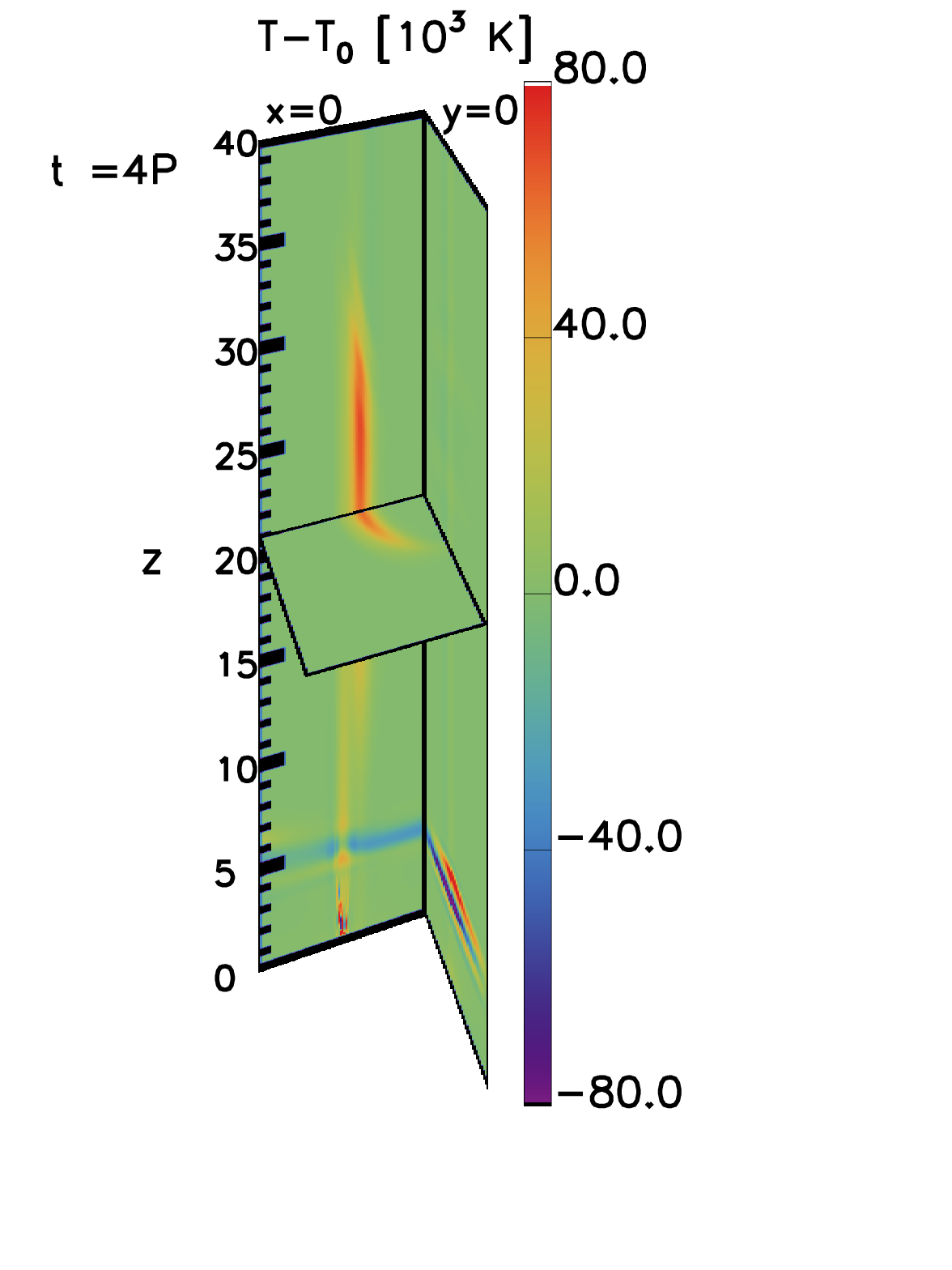} % printer
\caption{3D cuts of the MHD simulation where show maps of
density contrast $\left(\rho(t)-\rho(0)\right)/\rho(0)$ (left-hand side column),
$v_x$ (centre column),
and temperature change $\left(T(t)-T(0)\right)$ (right-hand side column)
on the $x=0$ and $y=0$ plane
and the horizontal plane at the $z=(t-P)V_{A0}$ coordinate at
$t=1$ $P$ (upper row) and $t=4$ $P$ (lower row)}
\label{3devol}
\end{figure}

The evolution of the MHD simulation
allows us to analyse how the process of 
mode coupling and phase mixing leads to the deposition 
of thermal energy in the system.
As soon as the driver sets in,
a wavetrain propagates into the domain
and we illustrate the evolution that follows in 
Fig.\ref{3devol}, where we show maps of
density contrast $(\rho(t)-\rho(0))/\rho(0)$,
$x$-component of the velocity and
temperature of the plasma at $t=P$ and $t=4$ $P$
in a projection of the 3D simulation box where
we cut two vertical planes at $x=0$ and $y=0$
and a horizontal plane at the $z$ coordinates of the 
trail of the wavetrain propagating at the Alfv\'en speed 
of the interior region.

The evolution described by the $y=0$ plane is the expected evolution 
of an oscillation of the flux tube along the $x$ direction 
where we see weak compression and rarefaction according to the phase of the driver,
alternate positive and negative $v_x$ regions and no significant 
temperature change. Additionally,
an acoustic mode follows the propagation of the transverse wave
and leads to a visible compression and rarefaction at $z=5$ $Mm$ at $t=4P$.

The evolution on the plane $x=0$ is of greater interest
for the present work.
The velocity patterns are in line with what was already found and thoroughly 
analysed by \citet{Pascoe2010} where the mode coupling and phase mixing
lead to the concentration of velocity structures in the boundary shell.
The amplitude of the driver (10\% of the Alfv\'en speed)
leads to weakly non-linear dynamics.

As the purpose of the work is to identify
the capacity of these phenomena to convert
the magnetic and kinetic energy concentrated in the boundary shell 
into plasma heating we address the temperature change
in the boundary shell at $t=4P$ (Fig.\ref{3devol} righ-hand side column).
We find that
a region of increased temperature along the vertical direction inside the boundary shell is formed.
This region has slightly variable width and it extends over about $45^{\circ}$
on the $xy$ plane as we see in the horizontal cut at $t=4$ $P$.
In this simulation the heating is of the order to $5\times10^4$ $K$,
going up to $7\times10^4$ $K$ in some regions.

\begin{figure}
\centering
\includegraphics[scale=0.30]{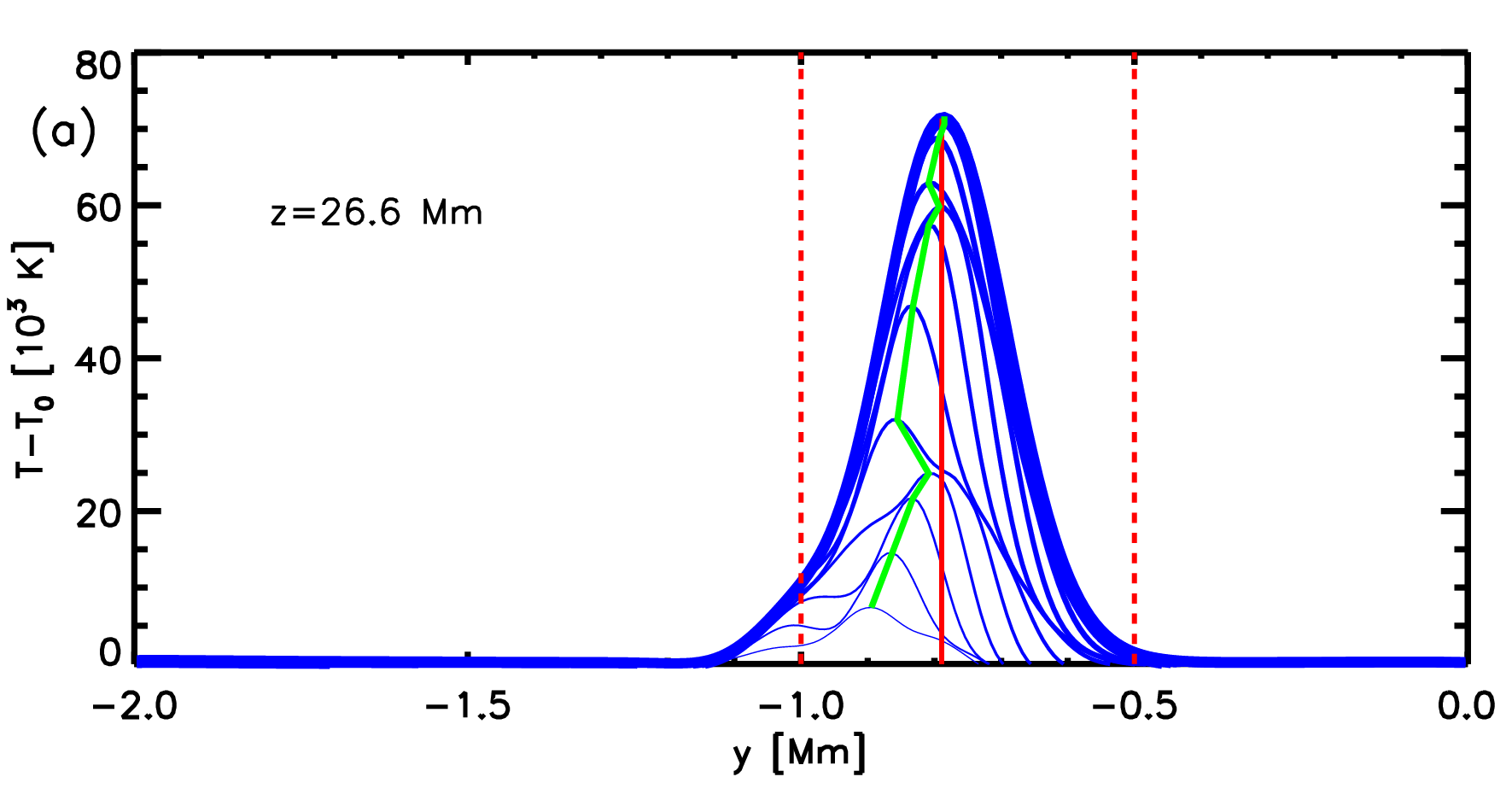} % printer
\includegraphics[scale=0.19]{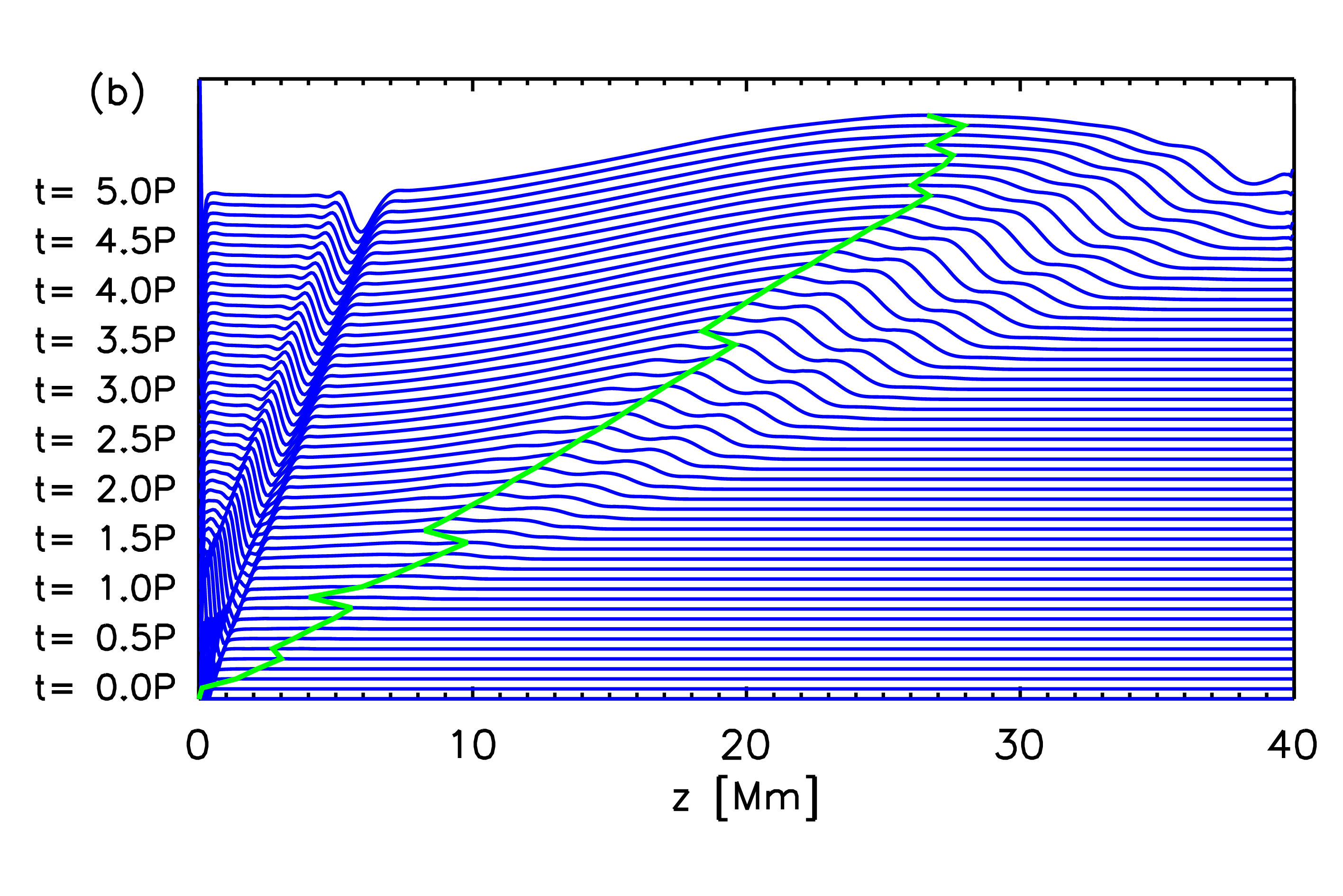} % printer
\caption{Temperature difference $\left(T(t)-T(0)\right)$ cuts.
(a) cuts at $x=0$ and $z=26.6$ $Mm$ (where the maximum temperature is found)
between $t=3.2$ $P$ (thinnest line) and $t=5$ $P$ (thickest line).
The vertical red dashed lines are the borders of the boundary shell and the red lines 
are placed at the location of the maximum gradient of Alfv\'en speed at each time.
(b) staggered cuts along z direction on the $x=0$ at the initial location of maximum gradient of Alfv\'en speed.
In both plots the green line follows the location of the maximum of temperature increase.}
\label{tempcuts}
\end{figure}
Fig.\ref{tempcuts}a shows the temperature increase profile at $z=26.60$ $Mm$
from $t=3.2$ $P$ to $t=5$ $P$ with a cadence of $0.1$ $P$ (from thinnest to thickest line)
where the red vertical line markes the position of the maximum of gradient of Alfv\'en speed.
The temperature starts to increase in the boundary shell near its border with the exterior region,
where the kinetic energy is higher.
Then the temperature increase profile maximum drifts towards the centre of the boundary shell
and stops at the position
where the gradient of Alfv\'en speed is maximum and from there it develops
in a temperature increase profile centred at that position.
The green line in Fig.\ref{tempcuts}a follows the maximum of the temperature increase
that gets closer to the centre of the boundary shell in three approaching segments,
each determined by the arrival of one of the velocity peaks induced by the driver.
The location of the maximum gradient of the Alfv\'en speed does not change over this time frame.
Fig.\ref{tempcuts}b stacks cuts of the temperature increase along the $z$-direction
at different times every $0.1P$ at the $y$ coordinate where
the maximum of the gradient of the Alfv\'en speed is situated.
We notice that the bump in the temperature increase is located at higher z 
as the evolution progresses until $t=4$ $P$,
when it settles at about $z=27$ $Mm$.
At the same time, its magnitude increases as well.
At $t=5P$ the maximum temperature increase is $7\times10^4$ $K$.
It also should be noted that at each time the maximum temperature increase lags
the perturbation as is visible from the wave
that propagates from the origin towards the edge of the domain.

\begin{figure}
\centering
\includegraphics[scale=0.37,clip,viewport=0 0 680 510]{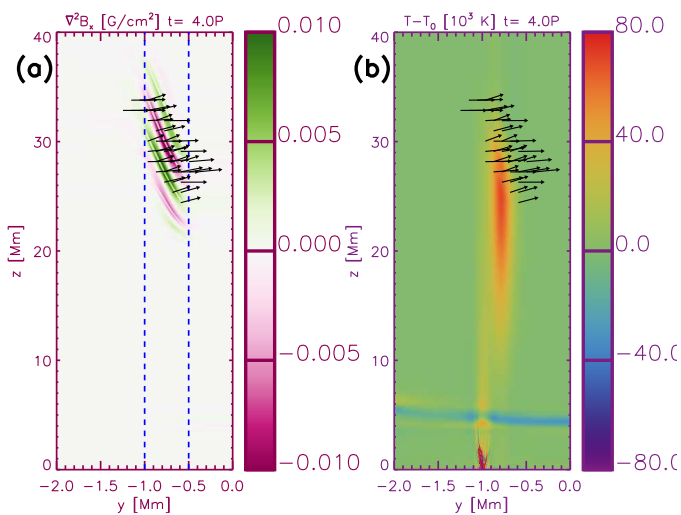} % printer
\caption{Maps on the $x=0$ plane of (a) $\nabla^2 B_x$,
and (b) temperature difference $\left(T(t)-T(0)\right)$
at $t=4$ $P$ with overplotted Poynting flux vectors where it is more intense.
In (a) the dashed blue lines define the borders of the boundary shell.} 
\label{db2temp}
\end{figure}
Similar conclusions can be justified and motivated from Fig.\ref{db2temp}
where we show maps of the quantity $\nabla^2 B_x$ (Fig.\ref{db2temp}a)
and temperature increase (Fig.\ref{db2temp}b) at $t=4P$
on the $x=0$ plane.
The black arrows represent the Poynting flux
(only where its intensity is above a threshold) and we choose $\nabla^2 B_x$
because it is some order of magnitude larger than $\nabla^2 B_y$ and $\nabla^2 B_z$.
As we are investigating a non-ideal heating mechanism we focus on $\nabla^2 B_x$
because where this term is large, the magnetic diffusivity operates and magnetic energy is converted into heating.
Where $\nabla^2 B_x$ is locally large, the Poynting vector 
shows a transfer of energy from the exterior to the boundary shell. This creates favourable conditions for the diffusivity term to act and thus, the conversion into thermal energy starts.
Therefore, the temperature increase becomes visible 
only after (or following) the transfer of energy to the boundary shell has occurred,
together with the formation of regions with high $\nabla^2 B_x$.

\begin{figure}
\centering
\includegraphics[scale=0.38]{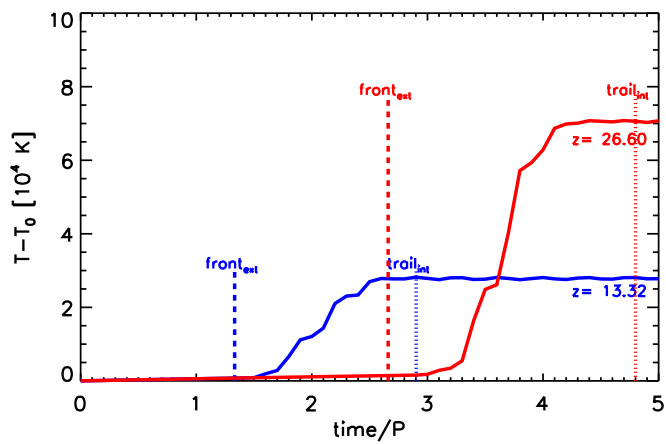} % printer
\caption{Temperature difference evolution of two points on the $x=0$ plane
at the location of the maximum gradient of the Alfv\'en speed 
at $z=26.60$ $Mm$ (red line - where the maximum temperature in the simulation is reached at $t=5$ $P$)
and at $z=13.32$ $Mm$ (blue line)}
\label{temppoint}
\end{figure}
Additionally, the heating takes place
over a bounded period of time from when the phase mixing 
starts to occur until the process of conversion of energy into thermal energy
is concluded.
Fig.\ref{temppoint} shows the temperature evolution of two points at
$z=13.32$ $Mm$ (blue line) and $z=26.60$ $Mm$ (red line)
at the $x=0$ plane at the location of maximum gradient of Alfv\'en speed.
The vertical lines show when a wave propagating from the lower boundary at the Alfv\'en
speed of the exterior region reaches the two points (fronts) and
when a wave propagating from the lower boundary when the driver ends
at the speed of the  Alfv\'en speed of the interior region (trails).
The temperature remains constant as long as no perturbation reaches
the point, it then steadily increases as the energy 
concentrated in the boundary shell is dissipated, and it finally remains constant again.
While this evolution is common for both points, 
the final temperature depends on the $z$ coordinate
because the amount of energy that can be converted into heating
depends on the intensity of the gradients of the magnetic field, which become steeper in z, the more out of phase adjacent propagating waves are.
At the same time because of the progressive damping of waves,
there is more kinetic energy available at lower z-coordinates
than higher up. In the present analysis
the time integral of the kinetic energy in the boundary shell that 
crosses the $z=13.32$ $Mm$ surface
is $20\%$ more than the one that crosses the $z=26.60$ $Mm$.

In order to further investigate how the energy is converted into heating,
we run a simulation where we set $\eta=0$ to
exclude the resistivity effects.
In MHD simulation terms this correspond to the adoption of numerical resistivity
which is the lowest resistivity value that a given MHD simulation can achieve.
The simulation with $\eta=0$ shows an evolution very similar to the one in Sec.\ref{simulation},
except that no significant temperature increase occurs due to phase mixing (less than $5\times10^3$ $K$).

\begin{figure}
\centering
\includegraphics[scale=0.30,clip,viewport=5 420 845 1270]{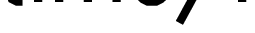} % printer
\caption{(a) Difference in energy in the boundary shell between time $t$ and $t=1\,P$:
magnetic (blue lines - $B_x$, magenta lines - $B_z$), kinetic (green lines), and thermal (red lines) energy
for the simulations with $\eta=10^9\eta_S$ and $\eta=0$.
(b) Difference in the energy in the boundary shell between the simulation with $\eta=10^9\eta_S$ and $\eta=0$
normalized to the thermal energy difference between the two simulations at each time.
Note that the horizontal axis in both panels starts at $t=1\,P$, i.e.~when the boundary driving has been terminated.}
\label{eta0compare}
\end{figure}
Fig.\ref{eta0compare}a shows the change of magnetic energy (associated with $B_x^2$ and $B_z^2$),
kinetic, and thermal energy in the boundary shell compared to time $t=P$ (when the driver has stopped) as a function of time in the two simulations.
The magnetic energy associated with $B_y^2$ remains negligible over the entire evolution.
The main difference is that the thermal energy significantly increases in the simulation where 
$\eta=10^9\eta_S$, while it shows only a very modest change in the simulation without resistivity.
This corresponds to a visible drop in magnetic ($B_x$) and kinetic energy in the simulation with resistivity,
when the driver has fed energy into the system.
In addition, the localised heating leads to an increase in the plasma pressure in the boundary shell
and hence (to conserve pressure balance), a decrease in $B_z$.
This decrease is clearly visible from the magenta lines in Fig.\ref{eta0compare}a.
The oscillations in the $B_z$-magnetic energy are due to the driver-induced transverse wave motions,
as the equilibrium magnetic field is not constant in the domain.
This analysis shows again that the non-ideal MHD terms are essential to convert
the energy concentrated in the boundary shell into heating.
In Fig.\ref{eta0compare}b we set equal to $1$ the difference of the thermal energy in the boundary shell at any given time
between the simulation with resistivity and the one without. The corresponding
difference in magnetic ($B_x$) and kinetic energy is about the same, at a value $-0.6$.
This result suggests that the kinetic and magnetic energy contribute to the same extend in 
providing the system with thermal energy as
expected for Alfv\'en waves and predicted in \citet{HeyvaertsPriest1983}.

\begin{figure}
\centering
\includegraphics[scale=0.30]{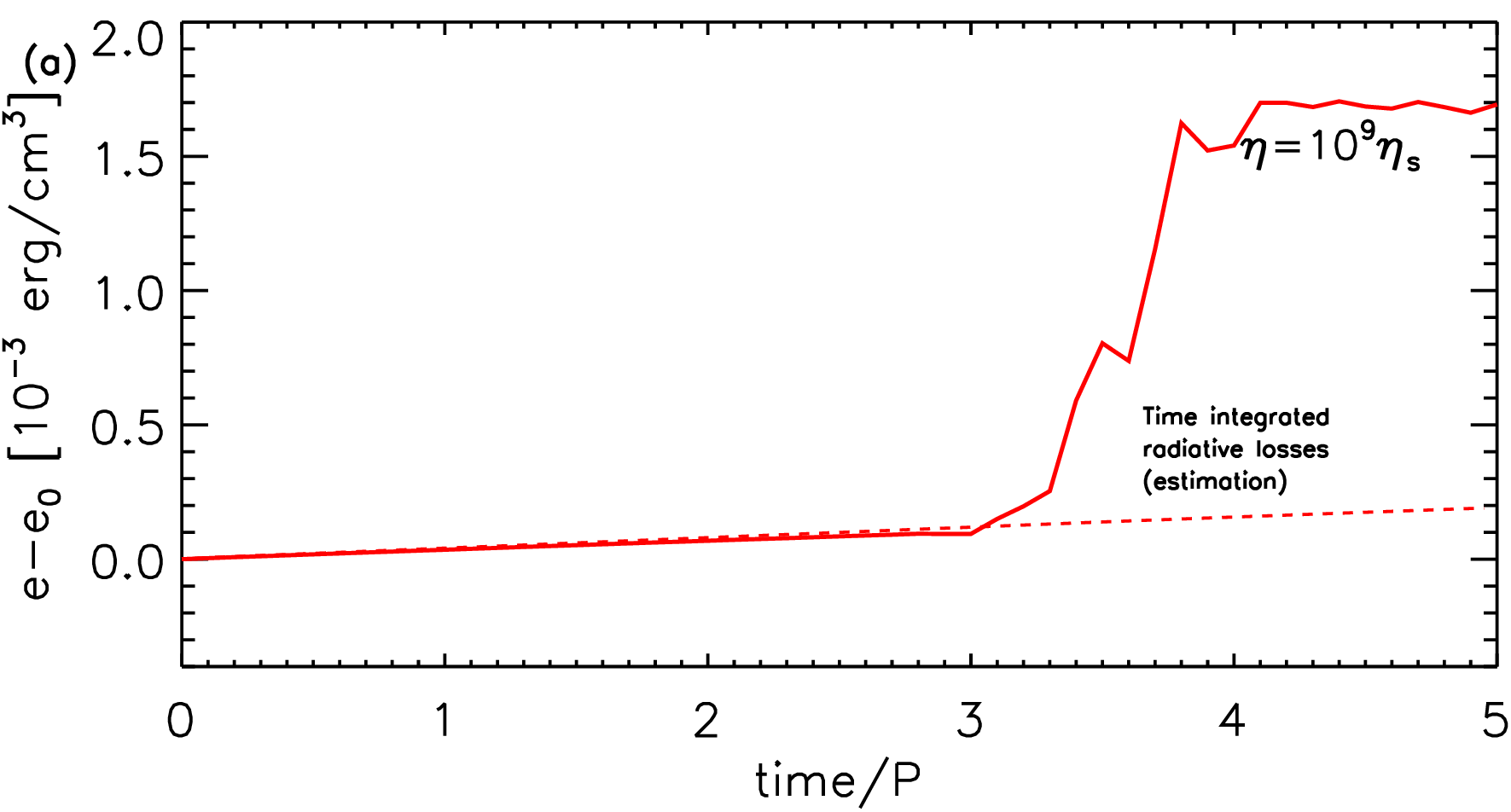} % printer
\includegraphics[scale=0.30]{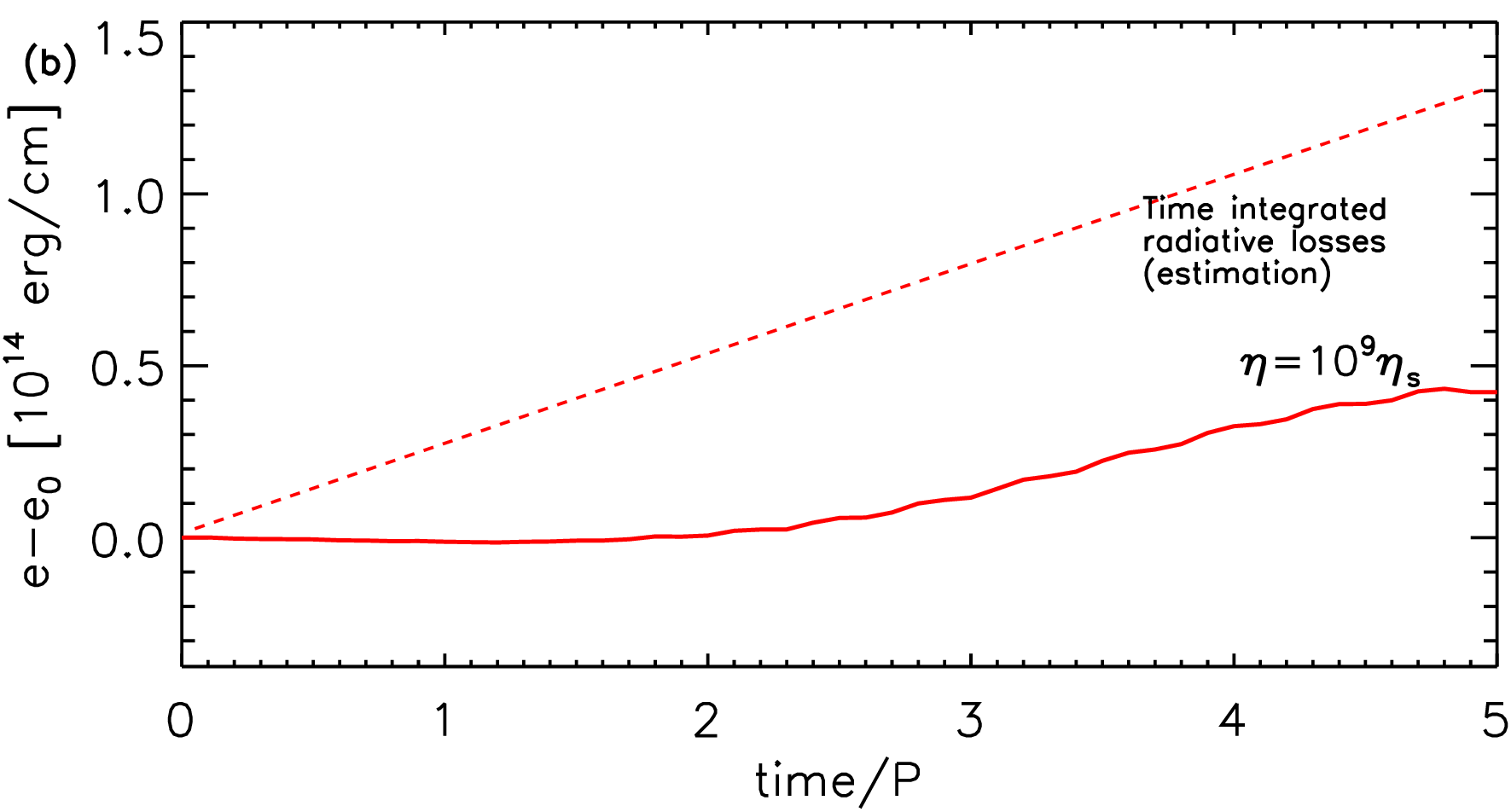} % printer
\caption{Thermal energy difference as a function of time
(a) in the same point as in Fig.\ref{temppoint} at $z=26.60$ $Mm$ and
(b) integrated in the entire $x=0$ plane.
In both plots the red dashed line is an estimation of the radiative losses for the interested region.}
\label{radlosses}
\end{figure}
However, the outstanding question is whether this energy input can match 
the radiative losses in order to answer to what extend this heating mechanism
can contribute to the coronal heating.
In Fig.\ref{radlosses}a we show the internal energy evolution in a plasma element located on the 
phase mixing shell at $z=26.60$ $Mm$
(same as red line in Fig.\ref{temppoint})
as a function of time compared with an estimation of the cumulated radiative losses 
following from the density and temperature evolution of the same plasma element.
We find that the energy deposition in a single plasma element can largely 
overcome the radiative losses.
However, the question is not if the heating mechanism can maintain coronal temperature in a single favourable location,
but if it can supply sufficient energy for entire coronal structures.
Fig.\ref{radlosses}b shows the result when the same analysis is carried out and integrated on the upper three quarters of the $x=0$ plane
of the present simulation
(we exclude the lower part to avoid spurious variations due to the lower boundary of the simulation and the propagation of the acoustic mode).
As the heating is localised in a narrow region near the phase mixing shell, the energy contribution to this extended domain
turns out to be insufficient to balance the energy lost by radiation of the plasma on the plane $x=0$.
This spatial domain is arbitrary and just to show the effect of the 
spatial limitation of the heating. Conditions would be even less favourable to maintain a coronal temperature in this regime if we considered a $3D$ domain instead.

\section{Parameters space investigation}
\label{parameterspaceinvestigation}

In Sec.\ref{simulation} we have described the mechanism 
that leads to the deposition of thermal energy in the boundary shell
due to the phase mixing of propagating Alfv\'en waves.
Our estimation of the thermal energy contribution from this mechanism
seems not to be conclusive whether it can definitely overcome the radiative losses.
In light of this, it is essential to address the role of the
boundary shell in this mechanism in order to assess how much 
the energy deposition can differ from what we have analysed so far.
To do so we run simulations where the 
boundary shell of the cylinder varies in width and 
where it has a more complex structure.

\subsection{Boundary shell width}
Using the setup explained in Sec.\ref{model}
we run two more simulations where we only change the width of the 
boundary shell by using
$l=0.75$ (wider boundary shell)
and $l=0.35$ (narrower boundary shell) 
with respect to Tab.\ref{tableparameters}.
These are analysed in combination with the reference simulation
with $l=0.5$ described in Sec.\ref{simulation} to investigate how the mode coupling and phase mixing are affected by Alfv\'en speed gradients. A narrower boundary shell leads to a steeper Alfv\'en speed profile (as all other parameters in the setup have remained unchanged), implying that phase mixing will become more efficient (i.e. the damping length will be shorter \citet{HeyvaertsPriest1983}). The mode coupling process which feeds kinetic and magnetic energy into the shell region also depends on the Alfv\'en speed profile in the shell region, but now a wider shell (less steep Alfv\'en speed gradient) leads to more efficient mode coupling \citep{Pascoe2010,Pascoe2012,Pascoe2013}.
Hence, the deposition of thermal energy is dependent on the combined efficiency of mode coupling and phase mixing
and how the resistivity effects interact with these mechanisms, so that
it is not a priori obvious which configuration will be most efficient.
The purpose of this experiment is to assess which effect dominates
the dynamics and how this influences the non-ideal effects in the simulation,
in particular the heating.

\begin{figure}
\centering
\includegraphics[scale=0.30,clip,viewport=010 80 367 675]{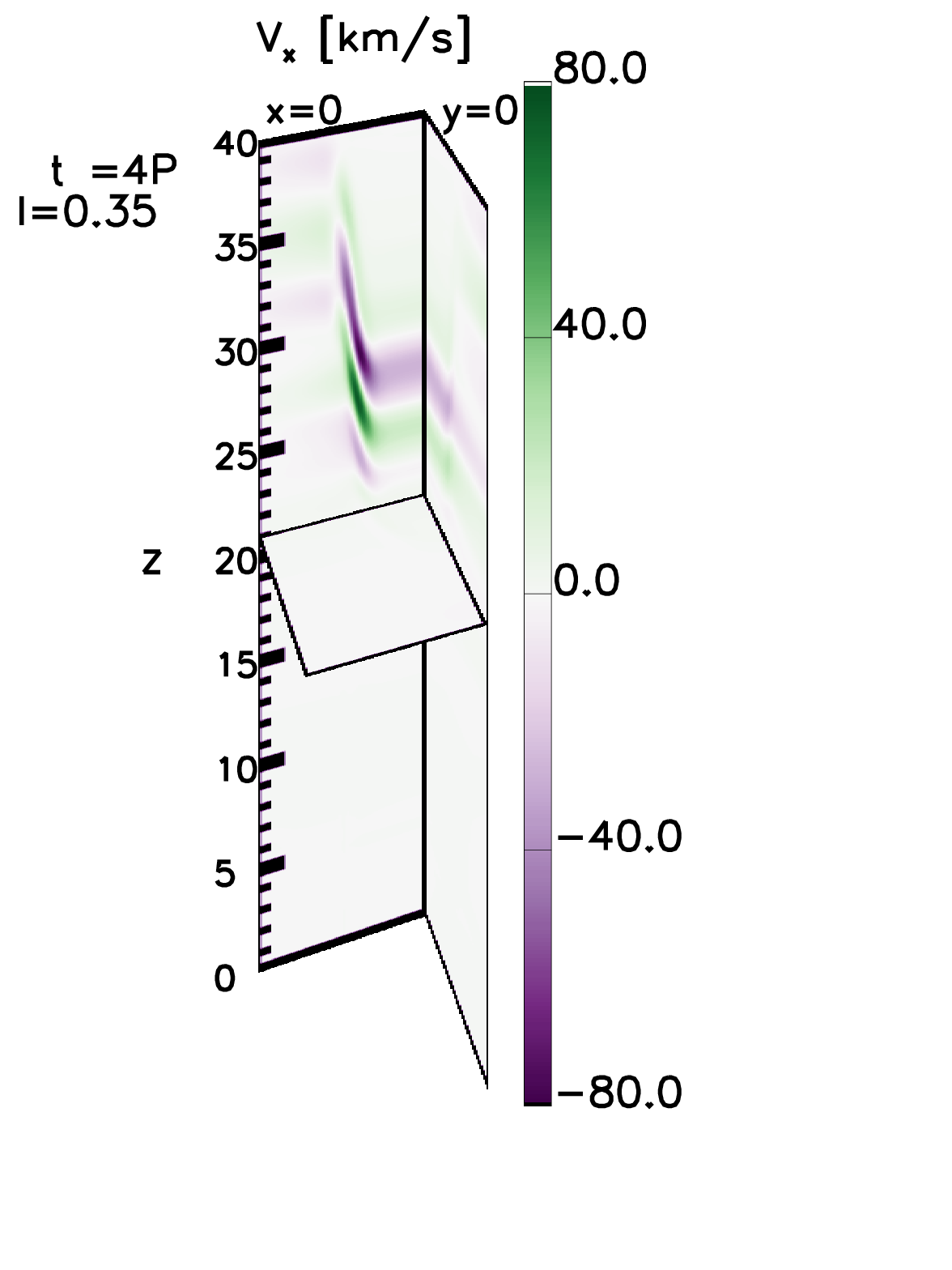}
\includegraphics[scale=0.30,clip,viewport=110 80 365 675]{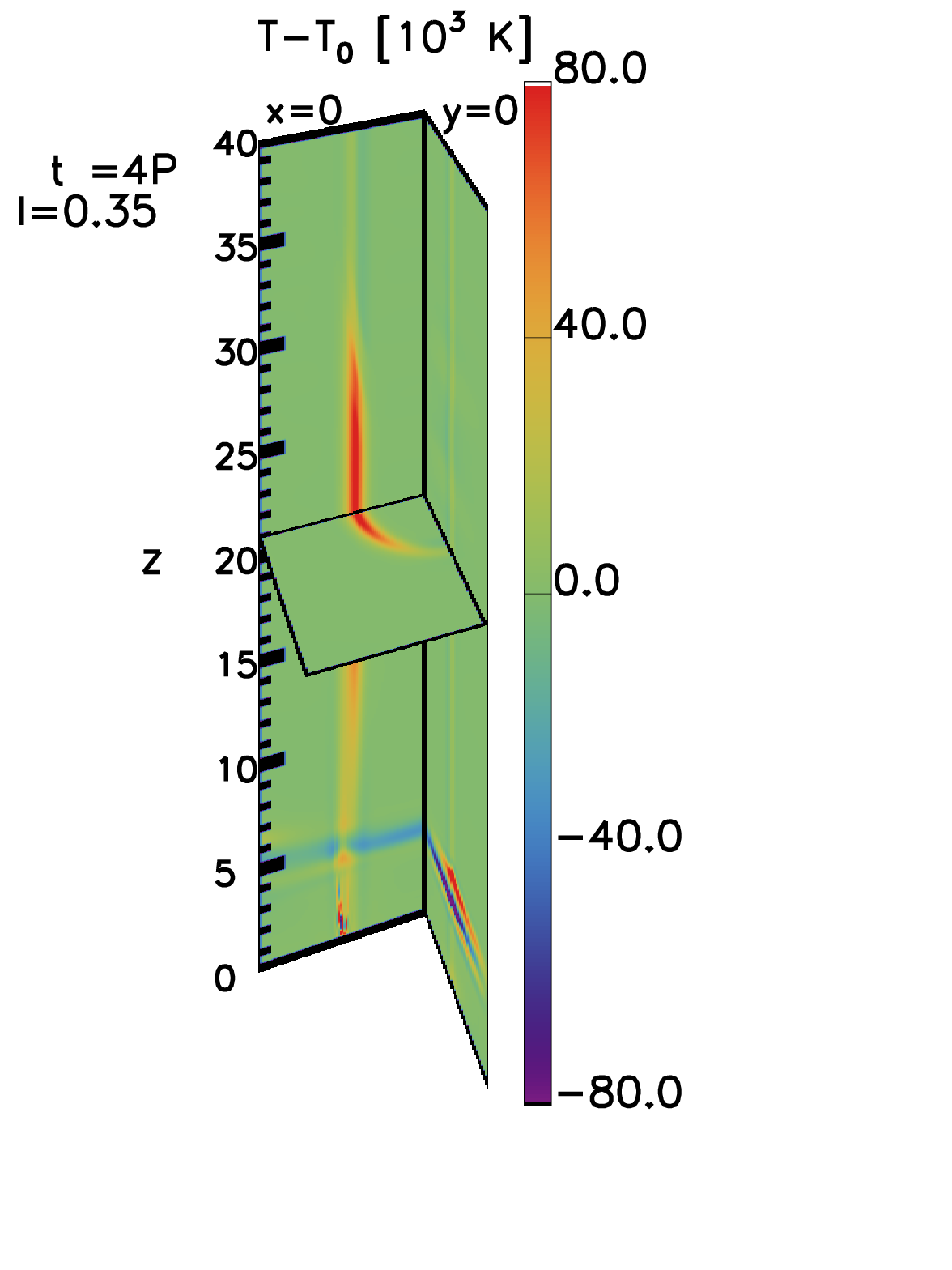}

\includegraphics[scale=0.30,clip,viewport=010 80 367 675]{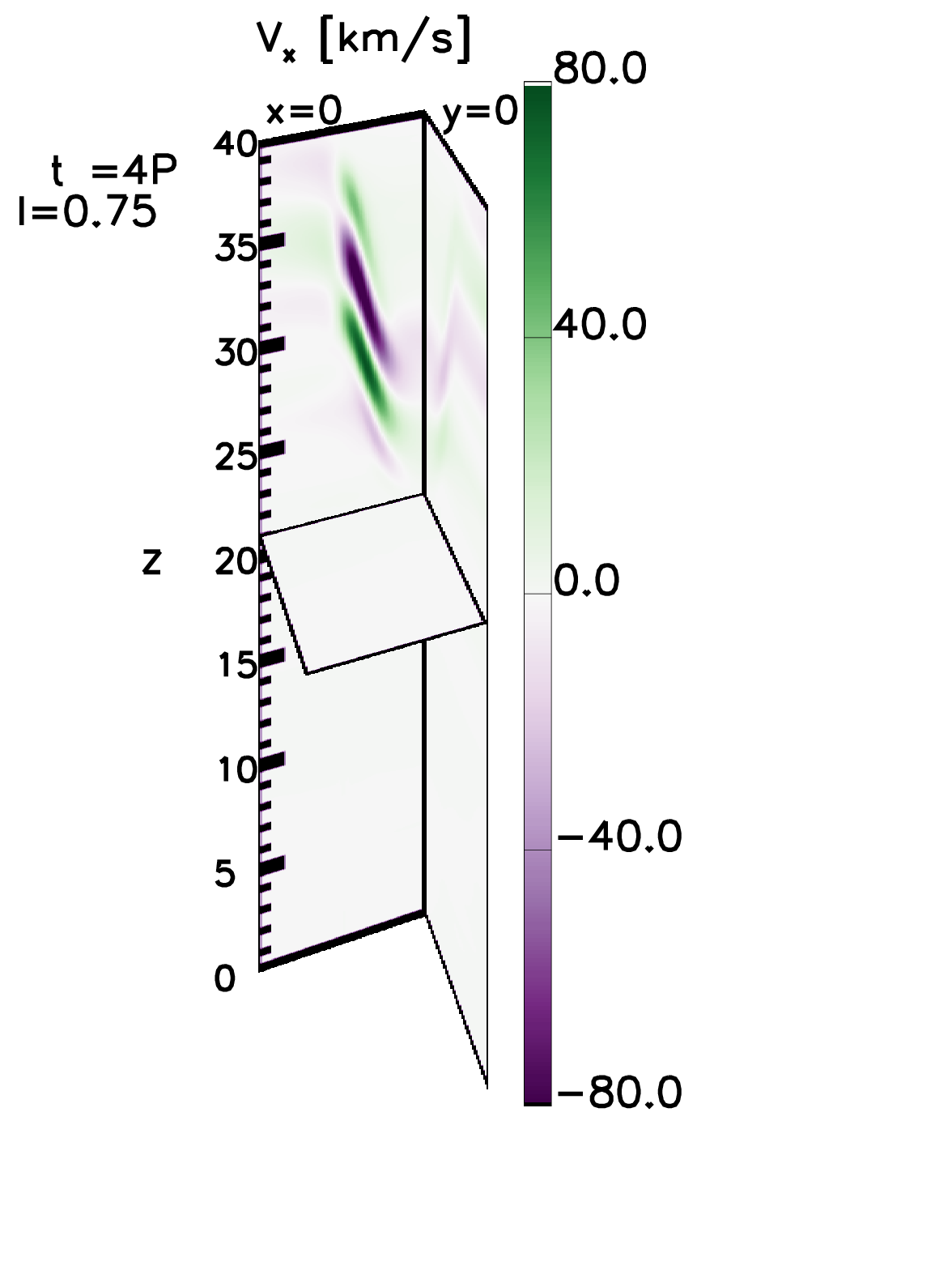}
\includegraphics[scale=0.30,clip,viewport=110 80 365 675]{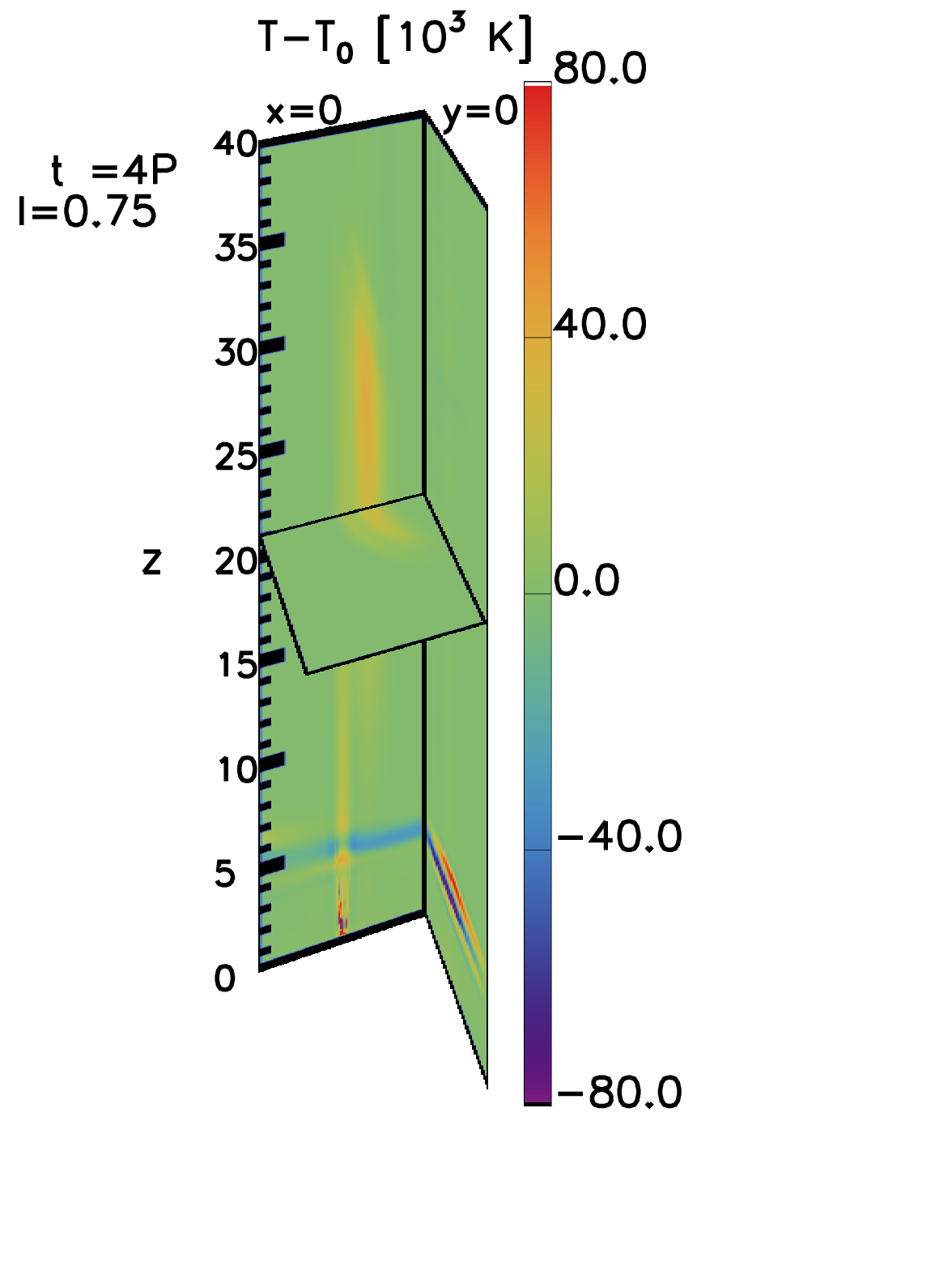}
\caption{3D cuts as in Fig.\ref{3devol} of $v_x$ left-hand side column and 
temperature difference, $\left(T(t)-T(0)\right)$, right-hand side column
at $t=4$ $P$ for the simulation with
$l=0.35$ (upper row) and with $l=0.75$ (lower row).}
\label{3devolbl}
\end{figure}
Fig.\ref{3devolbl} shows $3D$ cuts of $v_x$ and $T$ on the $x=0$ and $y=0$ planes at $t=4$ $P$
for the simulations with $l=0.35$ and $l=0.75$,
to be compared with the simulation with $l=0.5$ (Fig.\ref{3devol}).
The velocity phase-mixing patterns are visibly narrower in the simulation with $l=0.35$
and visibly wider in the simulation with $l=0.75$ as they simply 
fill the boundary shell.
Similar variations are found in the extension of the temperature increase region,
where the temperature increase is however dependent on the size of the boundary shell.
The region where the heating occurs is narrowest in the simulation with $l=0.35$
and the maximum temperature increase is $\Delta T\simeq10^5$ $K$, while
the simulation with $l=0.75$ show the largest heated region where $\Delta T\simeq5\times10^4$ $K$.

\begin{figure}
\centering
\includegraphics[scale=0.38]{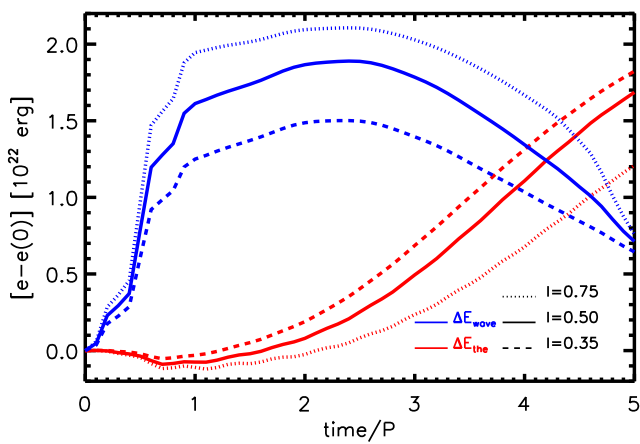} % printer
\caption{Evolution of the wave energy (blue lines) and thermal energy (red lines) in the loop boundary layer as a function of time for different widths of the boundary ($l=0.35$, dashed line;  $l=0.50$, continuous line; $l=0.75$, dotted line).}
\label{bldependency_energy_normvol}
\end{figure}
In order to assess which configuration (wide or narrow boundary shell)
leads to more efficient deposition of thermal energy,
we analyse the time evolution of the wave energy (kinetic $+$ magnetic energy associated to $B_x$)
and the thermal energy in the boundary shell (Fig.\ref{bldependency_energy_normvol}).
Initially ($t < 1\,P$), energy is injected into the domain (blue lines), including the boundary shell, by the driver.
This initial increase is followed by a second rise, where the mode coupling process transfers wave energy into the shell region.
However, phase mixing is already taking place at this stage, as is evident from the increase in thermal energy. 
Indeed, the thermal energy starts to rise before the wave energy reaches its maximum value.
Following this maximum, dissipation through phase mixing is clearly the dominant process in the shell region:
wave energy is converted into thermal energy at a faster rate than it is transferred into the layer by mode coupling.
Comparing the thermal energy curves (red lines),
it is clear that the increase in thermal energy is largest for the narrow boundary layer (dashed red line).
This is in agreement with Fig.\ref{3devolbl} which showed a larger increase in temperature in the boundary
layer at $t=4\,P$ for the simulation with the smallest boundary width ($l=0.35$).

\begin{figure}
\centering
\includegraphics[scale=0.38]{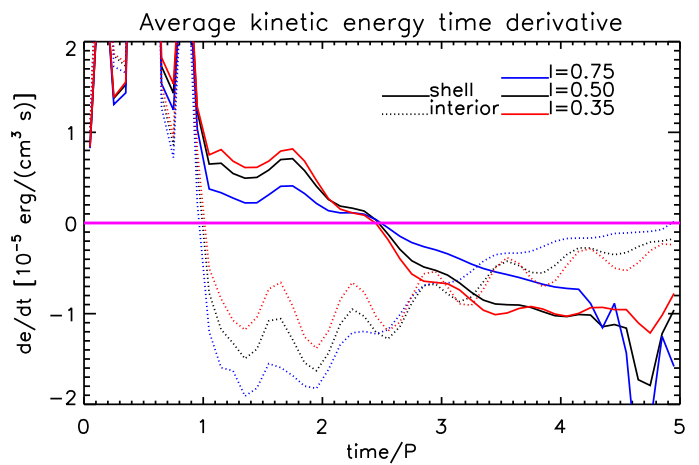}
\caption{Time evolution of the time derivative of the average kinetic energy
in the boundary shell (continuous lines) and in the interior region (dashed lines)
for the three simulations with $l=0.35$ (red lines), $l=0.50$ (black lines), $l=0.75$ (blue lines)}
\label{kinloopshell}
\end{figure}
In order to further describe how the width of the boundary shell affects the 
energy transfer to the boundary shell we show 
the time derivative of the average kinetic energy in the interior region and in the boundary shell 
in Fig.\ref{kinloopshell}.
We focus on the average kinetic energy because
i) it is the form of energy that is initially pumped into the system and more clearly drifts to the boundary shell
and ii) averaging over the domain we compensate for two effect:
first the narrower boundary shell simulation has more total kinetic energy because it has more plasma in the simulation box
and second the broader boundary shell simulation has more kinetic energy in the boundary shell because the shell is larger.
In Fig.\ref{kinloopshell} the time derivative is large and positive for $t<P$
while the driver is pumping kinetic energy 
into both the interior and boundary shell.
After $t=1\,P$ the average kinetic energy in the interior diminishes and it increases in the boundary shell.
The time derivative of the average kinetic energy in the interior is negative for all simulations,
and the simulation with $l=0.75$ is the one where it is minimum.
The time derivatives of the interior region and the boundary shell are opposite in sign 
and the total kinetic energy remains roughly constant, with variations of the order of 1\%.
This confirms that in this first phase the mode coupling dominates
the dynamics, that the kinetic energy is transferred from the interior to the boundary shell,
and that a broader boundary shell makes the energy transfer via mode coupling more efficient.
This phase continues until $t\sim2.5$ $P$ when the time derivative of the average kinetic energy in the boundary shell changes sign.
In this second phase, dissipation dominates and hence, the time derivative of the average kinetic energy in the boundary shell is negative for all the simulations.
In this regime the narrower boundary shell simulation is more efficient in dissipating energy as its time derivative is lower.
At the same time, the time derivative in the interior region approaches zero because there is less and less energy to dissipate,
without ever showing a regime change, as the waves keep damping in that region.
After $t=4$ $P$ the propagating waves interact with the upper boundary of the simulation box and the kinetic energy undergoes 
variations because part leaves the domain.

\begin{figure}
\centering
\includegraphics[scale=0.37]{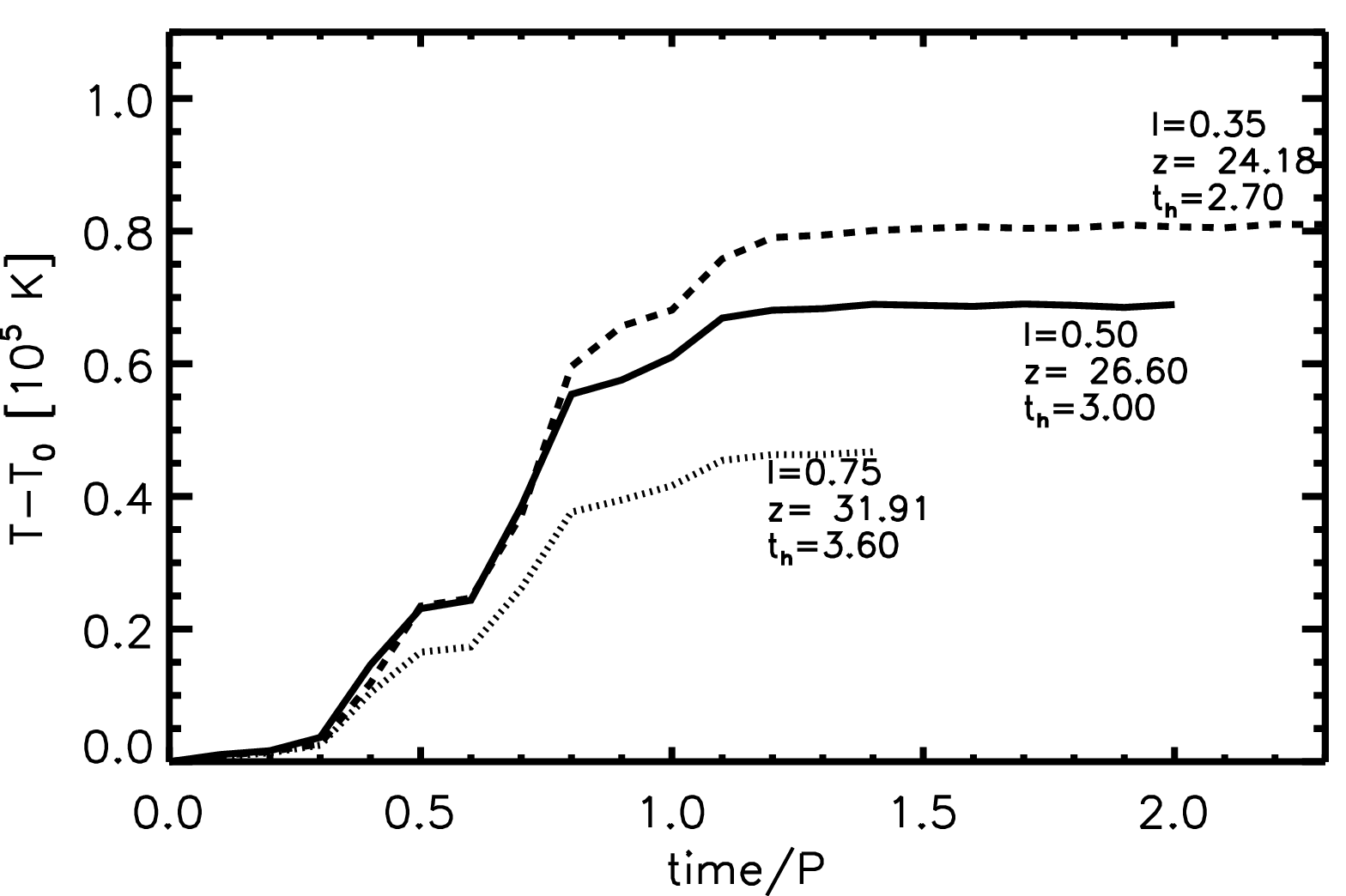} % printer
\caption{Temperature difference evolution of the points with highest temperature 
on the $x=0$ plane at the location of maximum gradient of Alfv\'en speed
in each simulation:
$l=0.75$ (dotted line), $l=0.50$ (continuous line), $l=0.35$ (dashed line).
Note that the time axes have been shifted so that
we plot the elapsed time from when the heating starts.
At the end of each line we report the $z$-coordinate of the point,
and the time at when the heating of the point starts.}
\label{pointtempsbl}
\end{figure}

Fig.\ref{pointtempsbl} shows the temperature evolution of the point 
on the $x=0$ plane that reaches the highest temperature at the end of the simulation
from the time when heating starts.
We find that the final temperature depends on the width of the boundary shell,
while the time scale over which thermal energy energy is deposited
is not dependent on the width of the boundary shell, 
as the time taken from the start of the temperature increase to the final temperature
is the same for all three simulations.
In all three simulations, the final temperature is reached in about one period.
Fig.\ref{pointtempsbl} also shows that the maximum of temperature is
reached at lower z (and earlier in time), as the boundary shell becomes narrower.

\subsection{Structure of the boundary shell}
\begin{figure}
\centering
\includegraphics[scale=0.37]{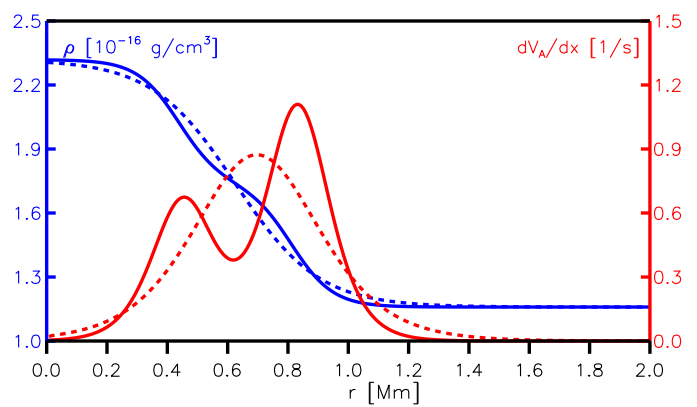} % printer
\caption{Density (blue lines) and gradient of Alfv\'en speed (red lines)
across the cylinder for the simulation with a smooth boundary shell (dashed lines)
and the simulation with a two steps boundary shell (continuous lines).}
\label{initdva2bl}
\end{figure}

The boundary shell structure we have investigated so far
is relatively simple,
where the density smoothly increases towards the interior region
and the gradient of the Alfv\'en speed
peaks near the centre of the boundary shell.
In order to address how the location of the heating
depends on the structure
of the boundary shell, we devise a different simulation where
we change the structure of the boundary into a profile with
two peaks in the gradient of the Alfv\'en speed (Fig.\ref{initdva2bl}).
In this simulation the boundary shell extends from
$b=0.25$ to $a=1$ $(l=0.75)$, and the density increases
in two steps alike Eq.\ref{densitylayer}, each extending for half of the boundary shell.
The density profile is given by:
\begin{equation}
\label{density2layer}
\displaystyle{\rho=0.5\left[\rho\left(\rho_e,\frac{\rho_2+\rho_i}{2},a,\frac{a+b}{2}\right)
                           +\rho\left(\frac{\rho_2+\rho_i}{2},\rho_i,\frac{a+b}{2},b\right)\right]}
\end{equation}
This density profile prescribes a boundary shell of width $0.75$ $Mm$
where the density in the interior and exterior region are the same as in the simulation with $l=0.75$.
The two gradient of Alfv\'en speed peaks are smaller and larger
than the equivalent profile for the simulation with a smooth boundary shell.

\begin{figure}
\centering
\includegraphics[scale=0.30,clip,viewport=010 80 367 675]{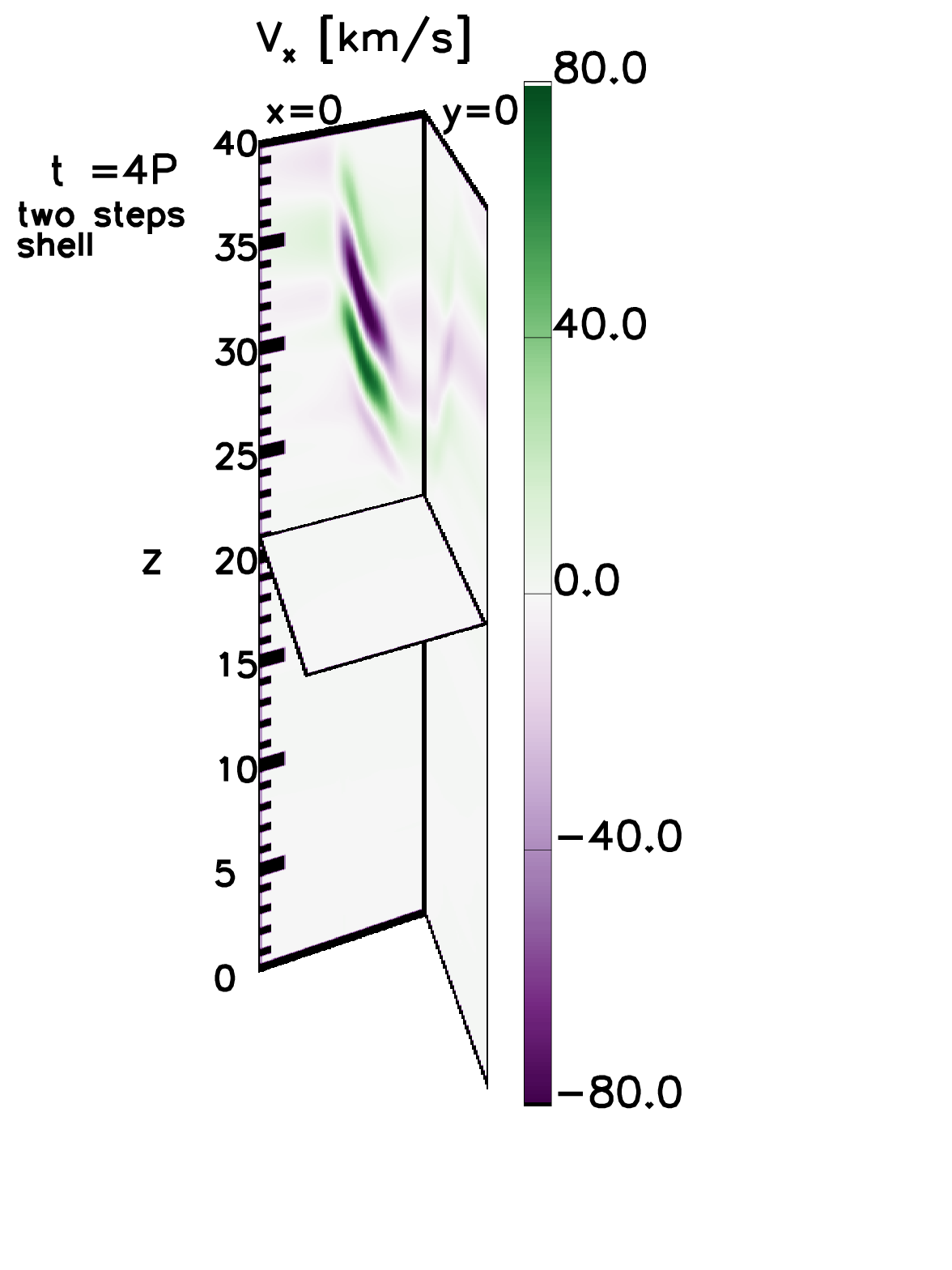}
\includegraphics[scale=0.30,clip,viewport=110 80 365 675]{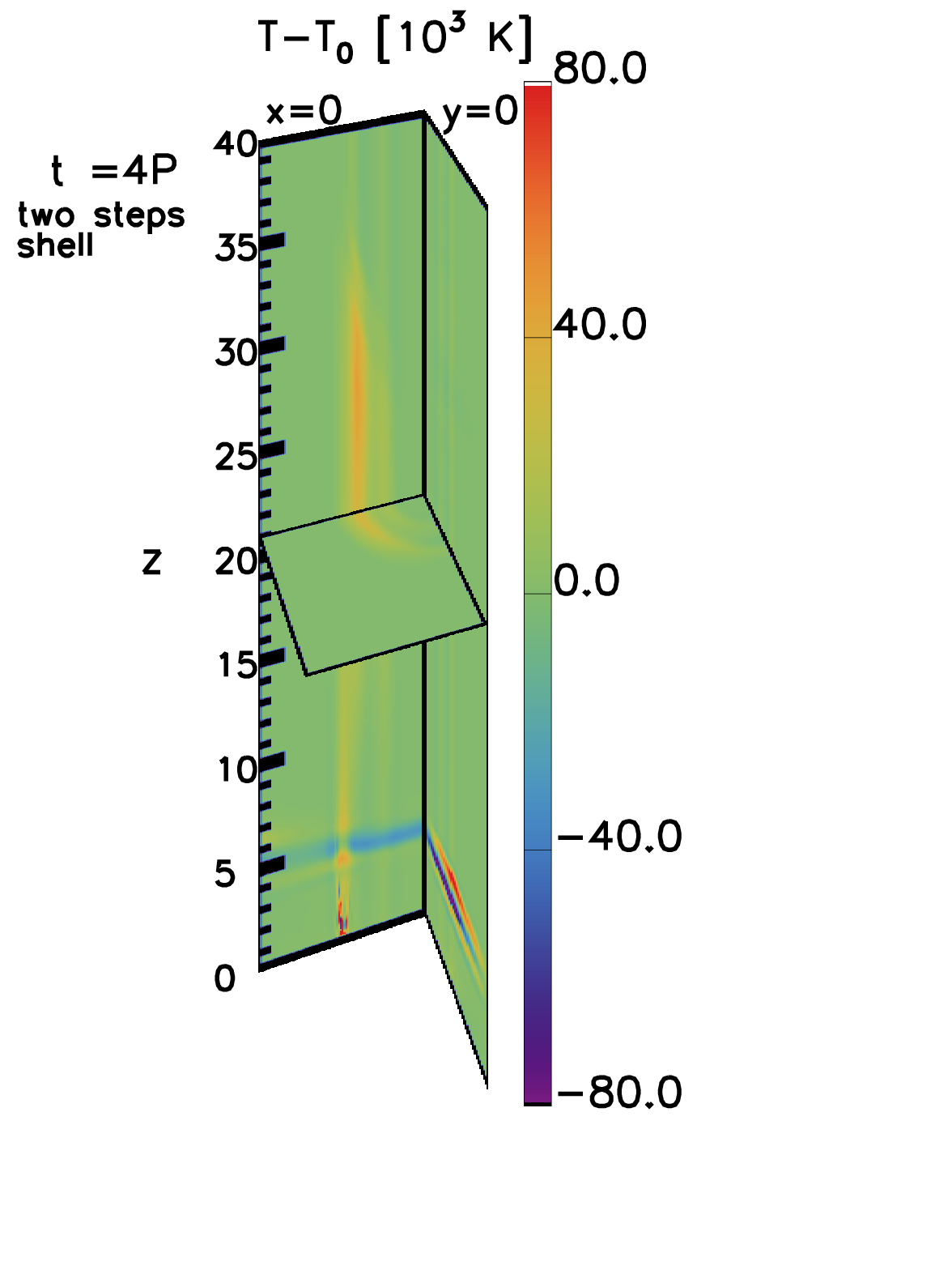}
\caption{3D cuts as in Fig.\ref{3devol} of $v_x$ left-hand side panel and 
tempertaure difference, $\left(T(t)-T(0)\right)$, right-hand side panel
at $t=4$ $P$ for the simulation with $l=0.75$ and a two steps boundary shell.}
\label{3devol2bl}
\end{figure}
Fig.\ref{3devol2bl} shows the evolution of the system at $t=4$ $P$
equivalent to Fig.\ref{3devolbl}
and the mode coupling and temperature increase patterns do not show big differences
with respect to the simulation with a smooth boundary profile.
We only find that the $v_x$ phase-mixing pattern becomes slightly more structured with a variable width 
and that the temperature increases at two different locations on the boundary shell, 
where the temperature increase is more signficant at the more external peak.

\begin{figure}
\centering
\includegraphics[scale=0.4]{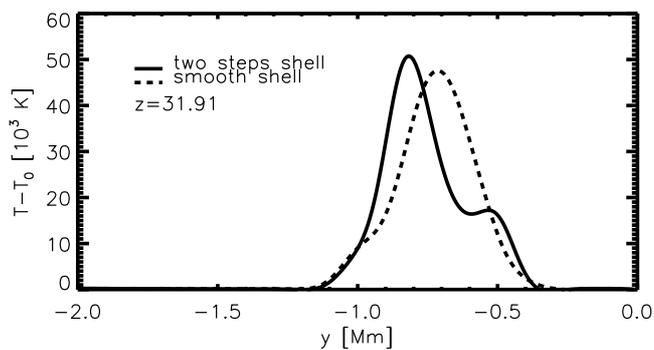}
\caption{Temperature difference, $\left(T(t)-T(0)\right)$, cut
across the cylinder on the $x=0$ plane at $z=31.91$ $Mm$
for the simulations with $l=0.75$ with 
the smooth boundary shell (dashed line)
and the two steps boundary shell (continuous line).}
\label{2blbl75tempcut}
\end{figure}
Fig.\ref{2blbl75tempcut} shows the temperature increase on the plane $x=0$
at $z=31.91$ $Mm$ at $t=5$ $P$ for the two simulations with boundary shell $l=0.75$ $Mm$.
We find that the temperature increase follows the structure of the boundary shell,
with as many temperature peaks as gradient of Alfv\'en speed peaks.
The peaks are also located in the same positions of the peaks of gradient of Alfv\'en speed
and the temperature increase is proportional to the intensity of the Alfv\'en speed gradient.

\begin{figure}
\centering
\includegraphics[scale=0.38]{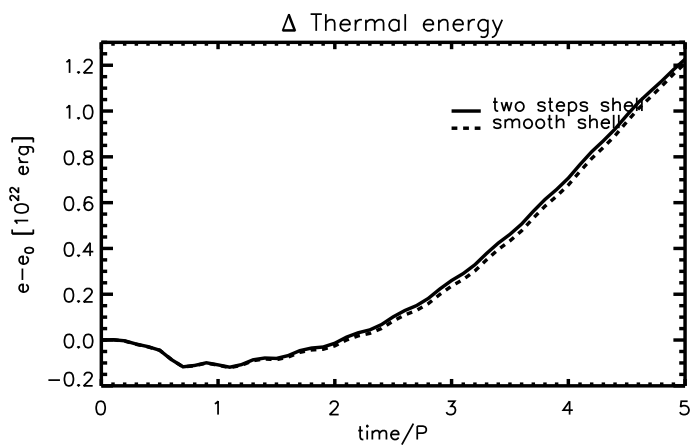} % printer
\caption{Thermal energy different between t and $t=0$ in the boundary shell
as a function of time for the simulation with $l=0.75$
and a smooth boundary shell (dashed line)
and a two steps boundary shell (continuous line).}
\label{ethe2bl}
\end{figure}
However, as shown in Fig.\ref{ethe2bl},
this does not lead to a visible change in the thermal energy deposited in the boundary shell
where in the simulation with a two step boundary shell only $~2\%$ more thermal energy is deposited in the boundary shell.

\subsection{Continuous driver}
\label{contdriver}
To conclude our investigation we analyse the evolution of
the system when a continuous driver (instead of a single pulse) is used to perturb the system.
This is a step towards a configuration where
the footpoints are continuously displaced by photospheric motion.
In this simulation we use all the parameters outlined in Sec.\ref{model},
with the only difference that we do not switch off the driver after one pulse, but
we let it continue.

\begin{figure}
\centering
\includegraphics[scale=0.30,clip,viewport=010 80 367 675]{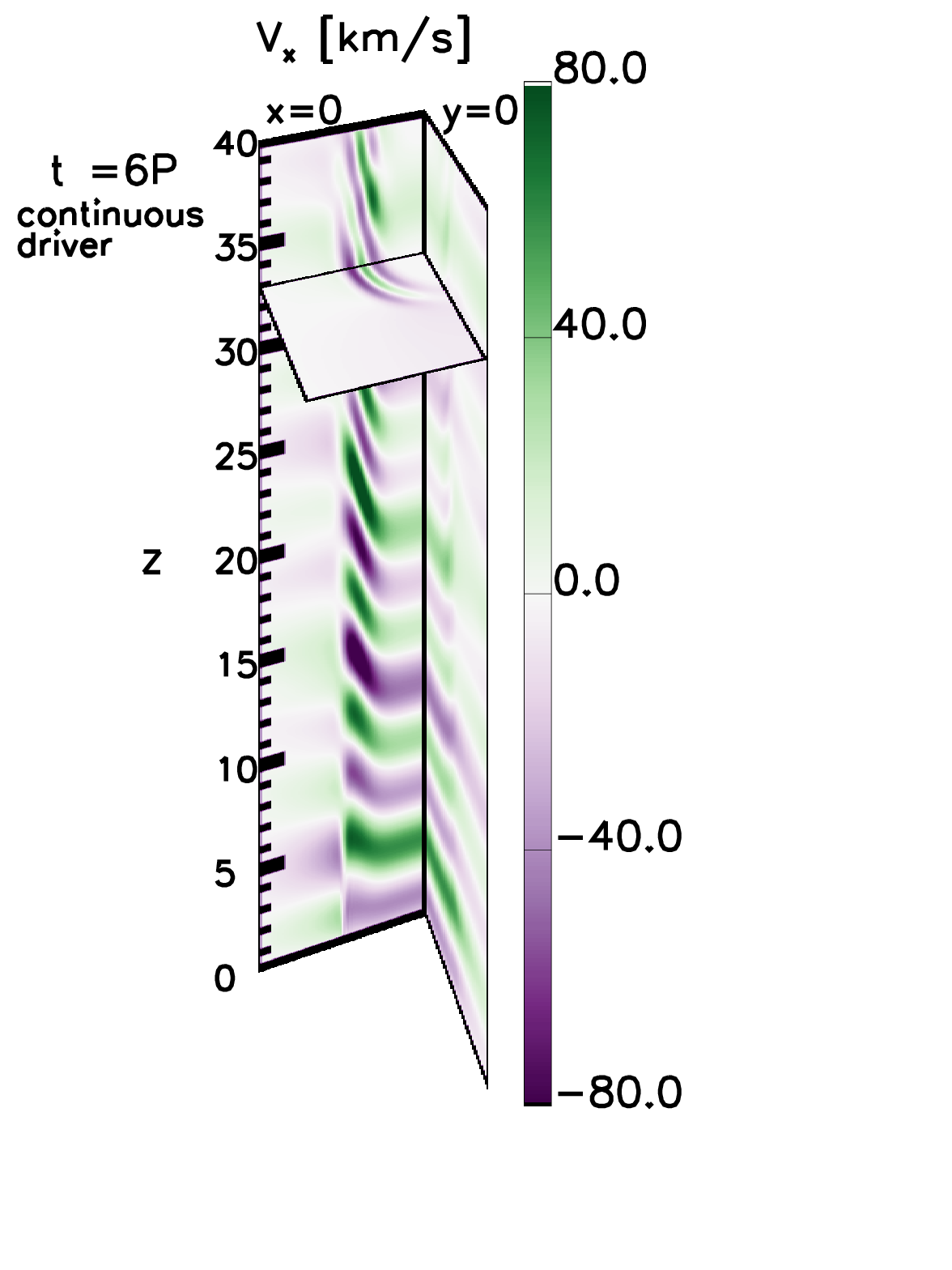}
\includegraphics[scale=0.30,clip,viewport=110 80 365 675]{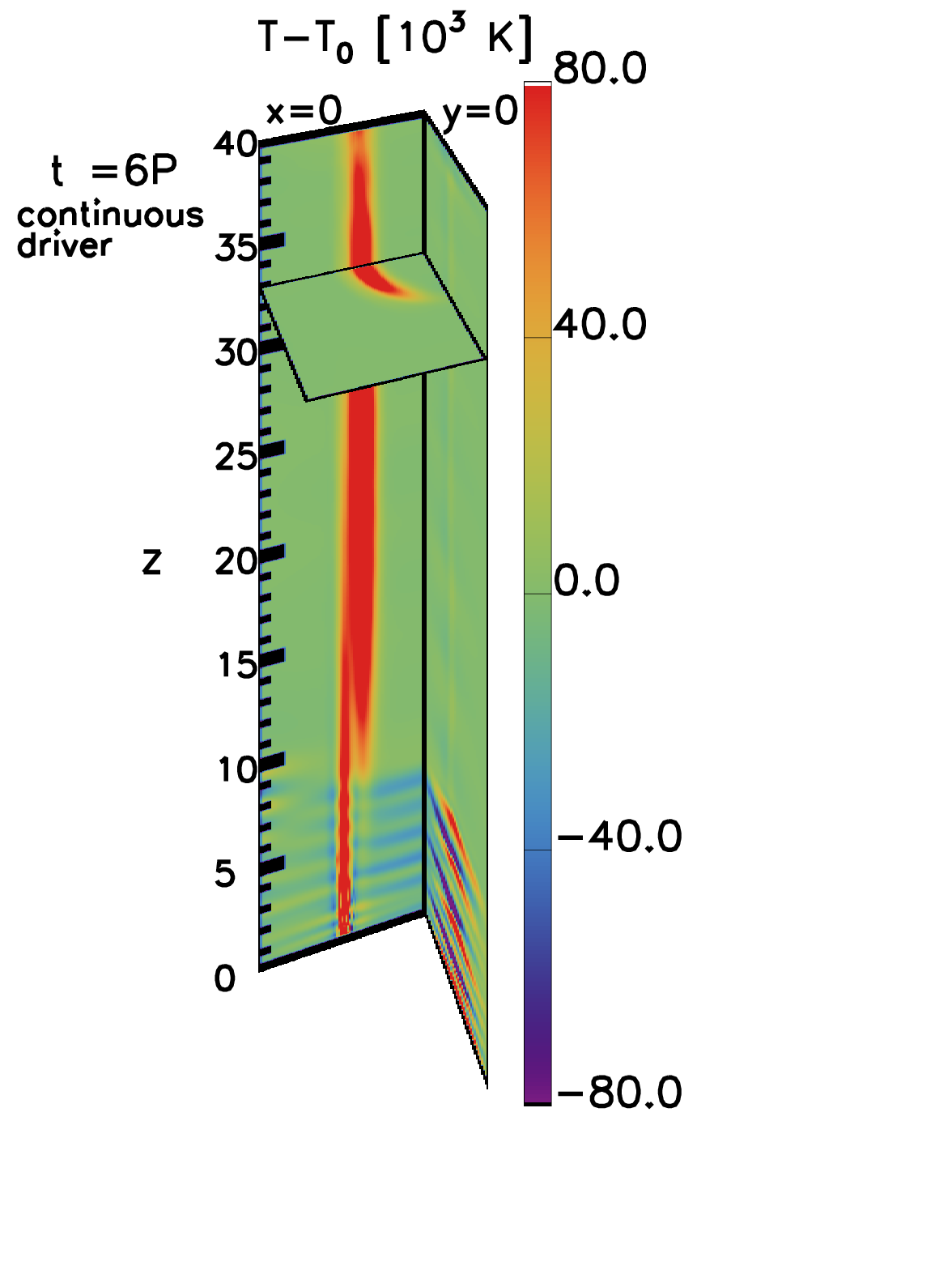}

\includegraphics[scale=0.30,clip,viewport=010 80 367 675]{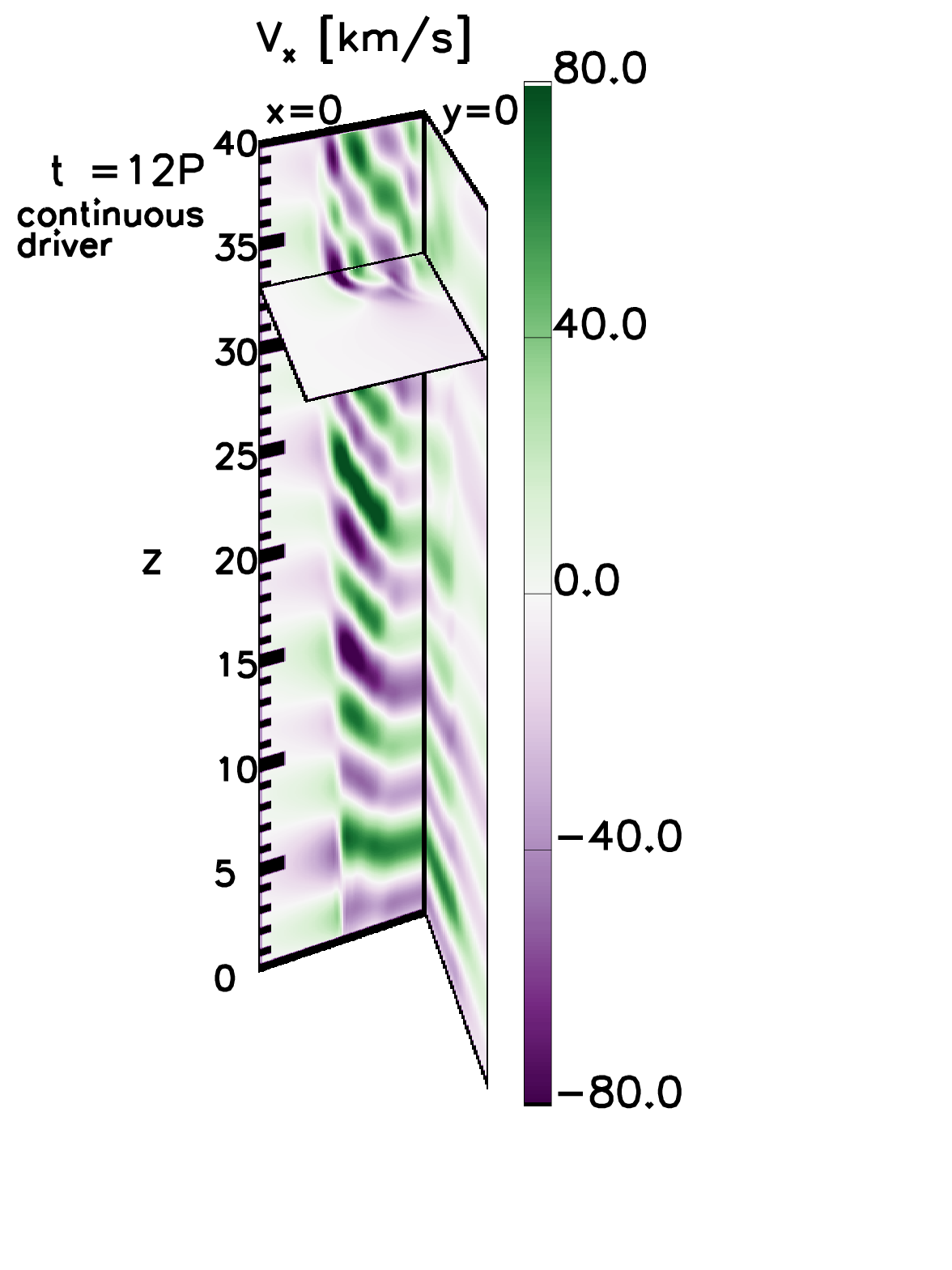}
\includegraphics[scale=0.30,clip,viewport=110 80 365 675]{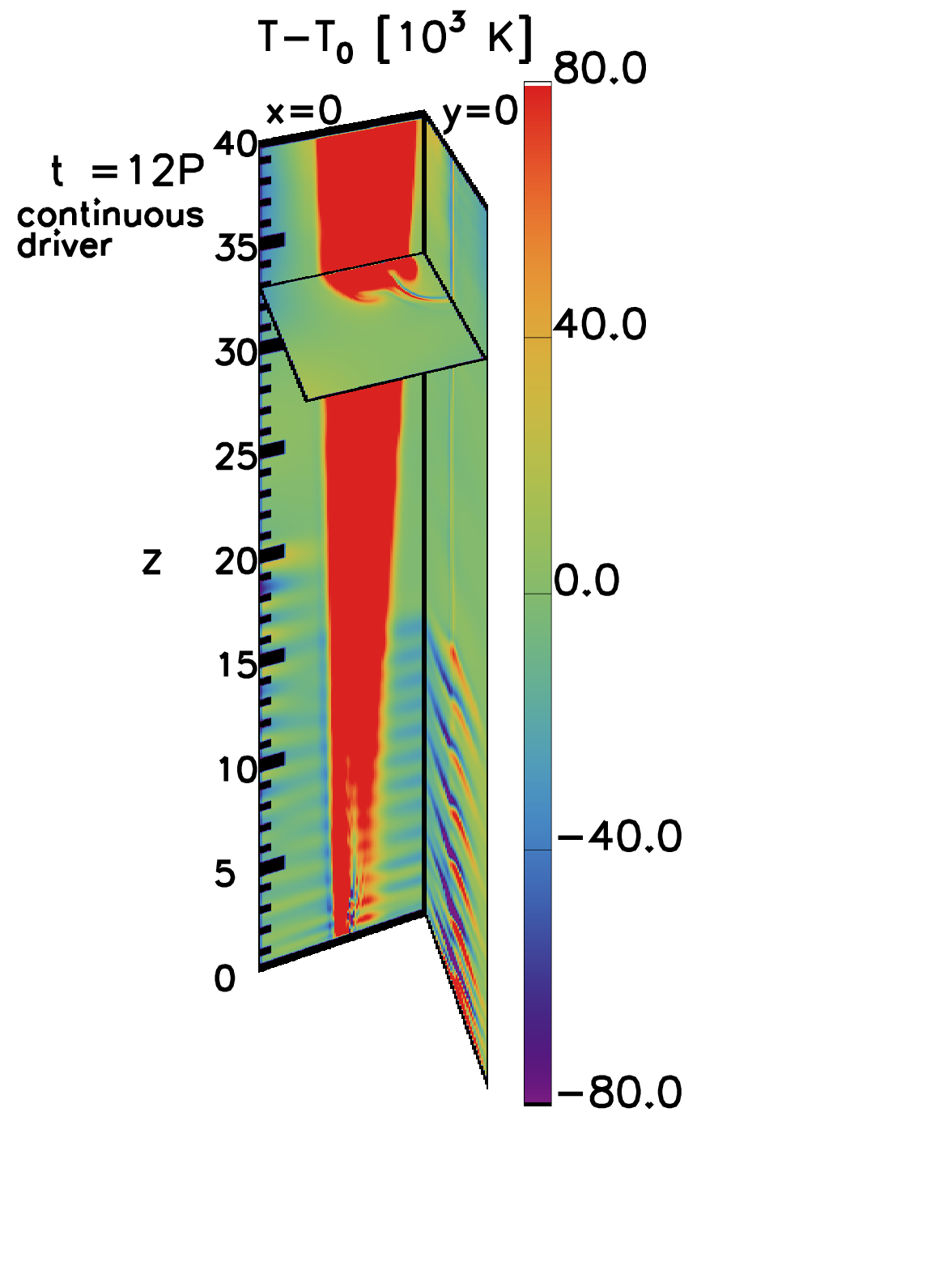}
\caption{3D cuts as in Fig.\ref{3devol} of $v_x$ left-hand side column and 
temperature difference, $\left(T(t)-T(0)\right)$, right-hand side column
at $t=6$ $P$ (upper row) and $t=12$ $P$ (lower row)
for the simulation with continuous driver.}
\label{3devol10T}
\end{figure}
Fig.\ref{3devol10T} shows the evolution of the system after 
$t=6$ $P$ and $t=12$ $P$.
The $v_x$ pattern shown on the plane $x=0$ indicates 
that the phase-mixing occurs in a similar way for each consecutive 
period of the driver as the same pattern visible in Fig.\ref{3devol}
repeats.
However, as each wave train encounters 
conditions increasingly departing from the initial condition,
the shape of the phase-mixing pattern becomes less regular and 
with more internal structuring.
The temperature increase pattern shows significant differences
with respect to our previous simulation,
as the plasma reaches higher temperatures
and the heated region broadens in time.

\begin{figure}
\centering
\includegraphics[scale=0.37]{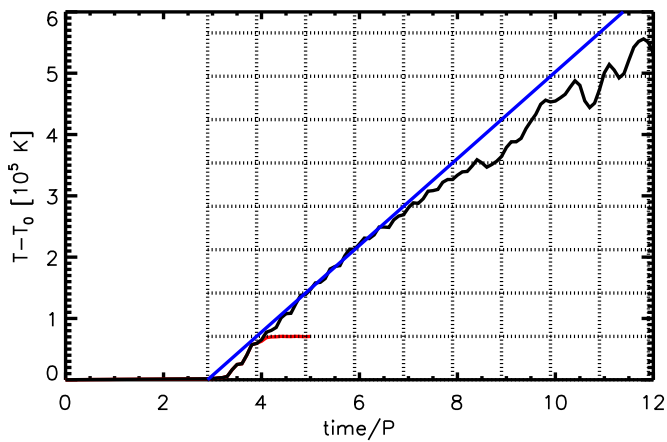} % printer
\caption{Temperature difference evolution of the point on the $x=0$ plane
at the location of the maximum gradient of the Alfv\'en speed 
at $z=26.60$ $Mm$ for the simulation with continuous driver (black continuous line),
compared to the same temperature evolution for the simulation with single pulse (red line).
The overplotted grid is horizontally spaced by $1$ $P$ and
vertically spaced by the final temperature increase in the single pulse simulation.
The blue straight line would be the temperature evolution is the 
same initial temperature increase is gained at each pulse.}
\label{pointtemps10T}
\end{figure}
Fig.\ref{pointtemps10T} shows the temperature increase evolution 
of a single plasma element located at the steepest Alfv\'en speed location
on the $x=0$ plane at $z=26.60$ $Mm$ 
and the overplotted grid is 
spaced one period horizontally and vertically by the temperature increase after one pulse.
The temperature increase starts after $t=3$ $P$,
and it increases mostly linearly until $t=7$ $P$,
when 4 pulses have crossed the plasma element.
After $t=7$ $P$, the temperature increase is no longer linear in time and
each pulse contributes with an increasingly smaller temperature rise.
At $t=12$ $P$ the temperature increase slightly exceeds $\sim5\times10^5$ $K$.

\begin{figure}
\centering
\includegraphics[scale=0.38]{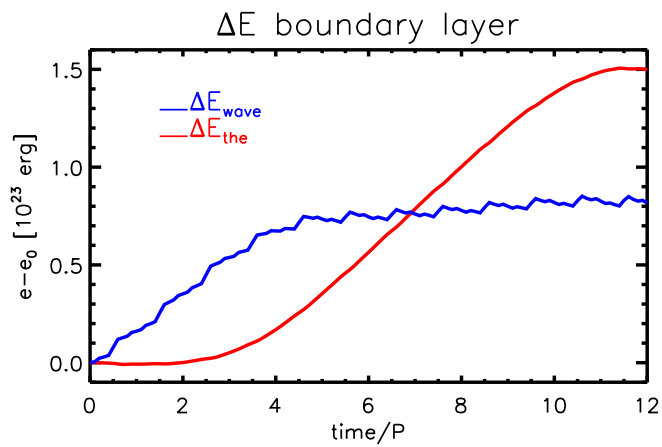} % printer
\caption{Difference in energy in the boundary shell between time $t$ and $t=0$
for the simulations with continuous driver as function of time:
wave energy (blue line) and thermal energy (red line)}.
\label{energylayer10T}
\end{figure}
Similar conclusions can be drawn from the evolution of energy in the boundary shell (Fig.\ref{energylayer10T}),
where we find that the thermal energy steadily increase until $t=11$ $P$.
After $t=2$ $P$, the thermal energy increases at the expense of the wave energy that enters the boundary shell.
However, the wave energy remains constant after $t=5$ $P$, when the entire z-extension of the domain is 
filled by the wave propagation.
After $t=11$ $P$, the thermal energy gets saturated.

\begin{figure}
\centering
\includegraphics[scale=0.3]{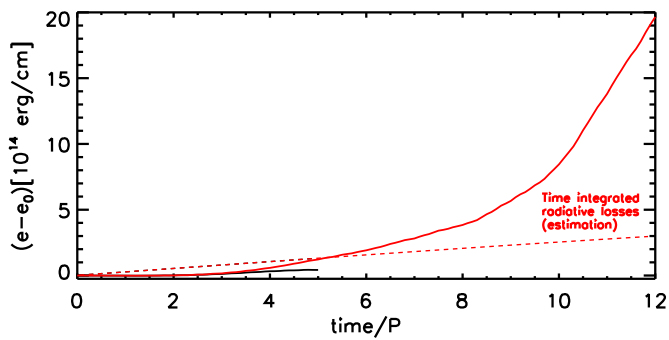} % printer
\caption{Thermal energy difference as a function of time integrated in the entire $x=0$ plane
for the simulation with continuous driver (red line) and for the simulation with single pulse (black line).
The red dashed line is an estimation of the radiative losses in the same region.}
\label{radlosses10T}
\end{figure}
Fig.\ref{radlosses10T} compares the thermal energy increase in the boundary shell at $x=0$
with an estimation of the radiative losses (as in Fig.\ref{radlosses}).
In this simulation the thermal energy deposition largely overcomes
the radiative losses because of the continuous injection of energy in the system.
It also should be noted that the increase in thermal energy is not linear and 
each pulse deposits an increasingly higher amount of thermal energy
because the deposition of energy drifts to lower and higher
z coordinates after the shells initially involved saturate.

\begin{figure}
\centering
\includegraphics[scale=0.18]{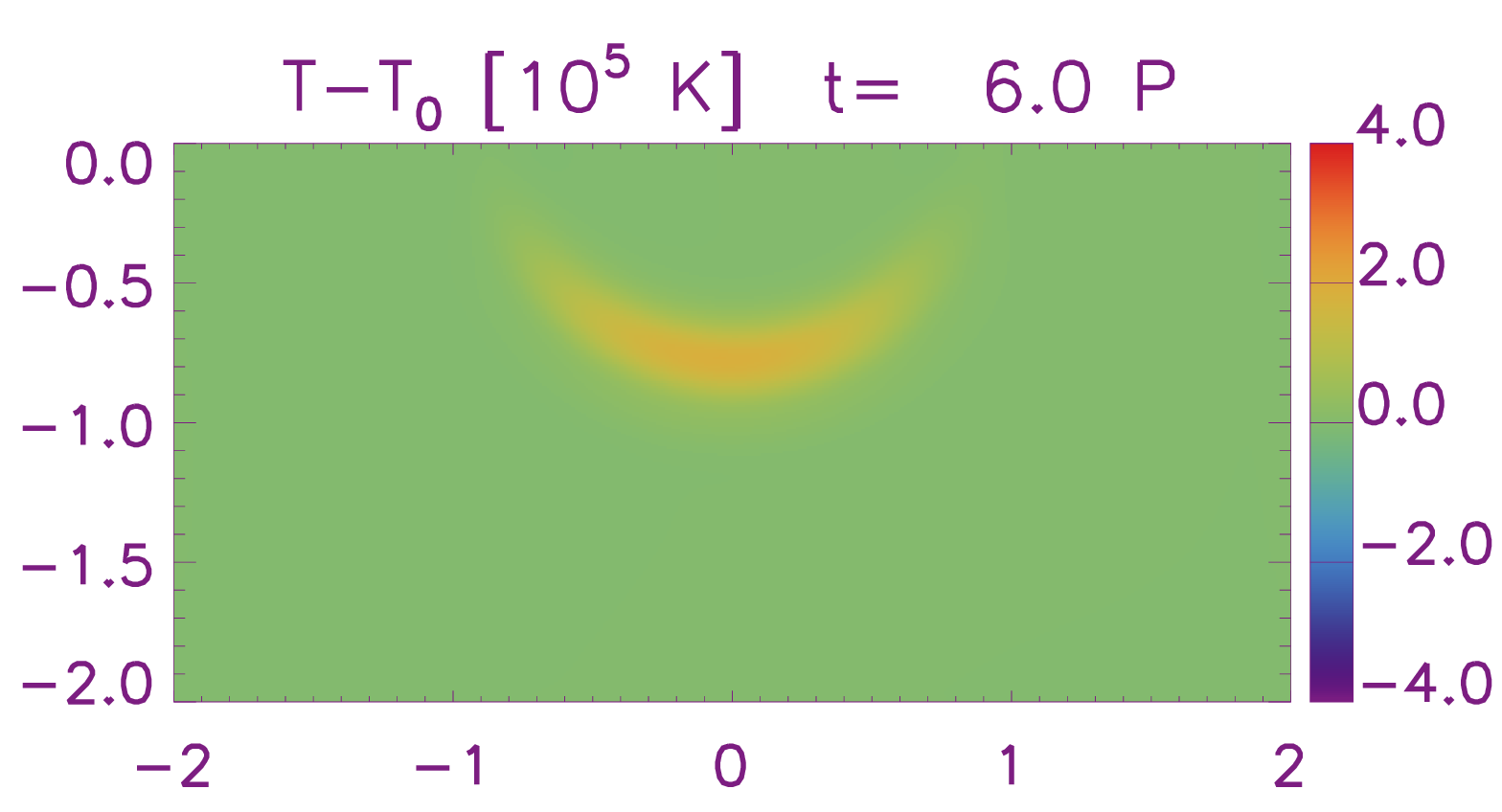} % printer
\includegraphics[scale=0.18]{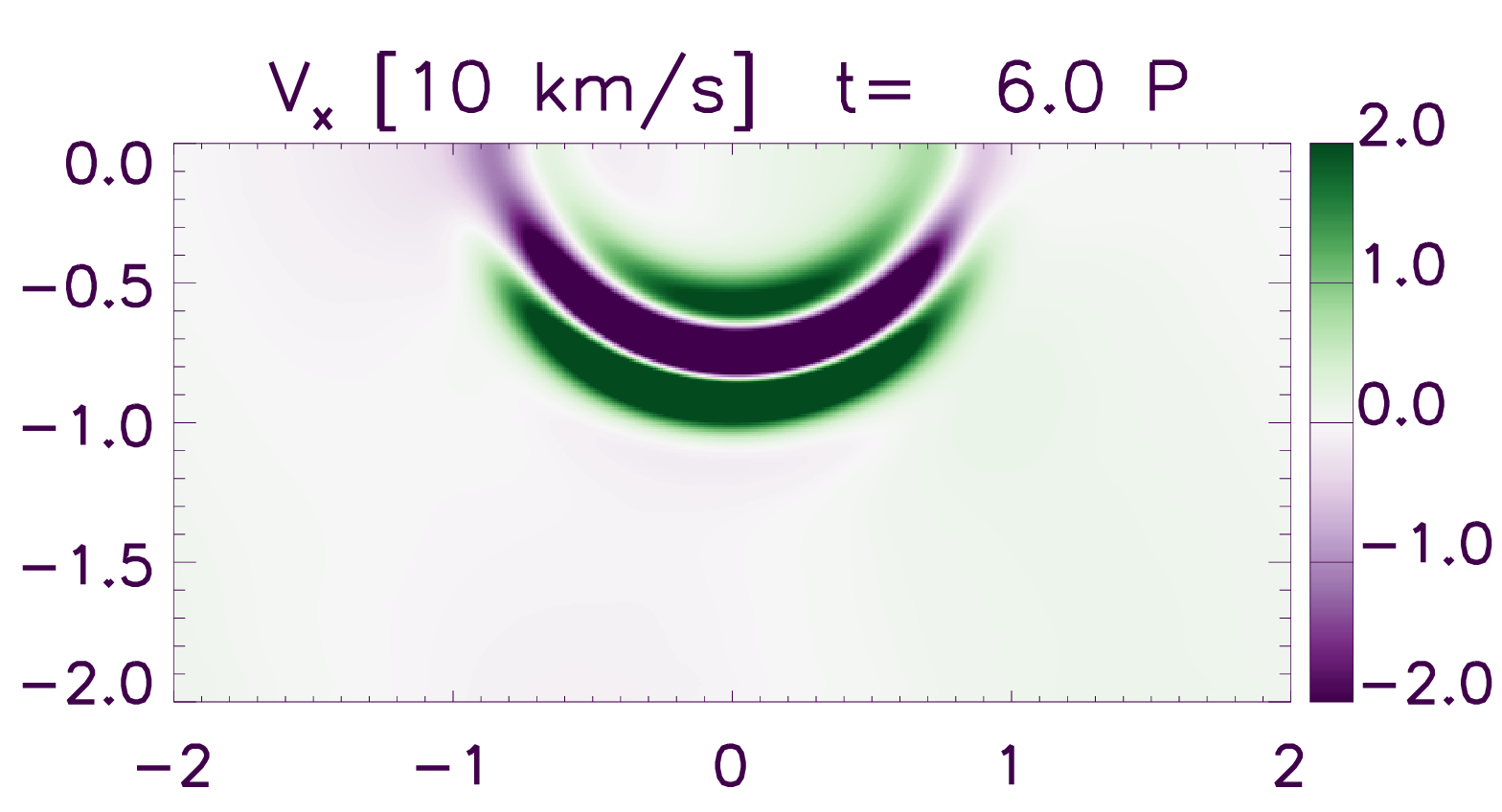}  % printer

\includegraphics[scale=0.18]{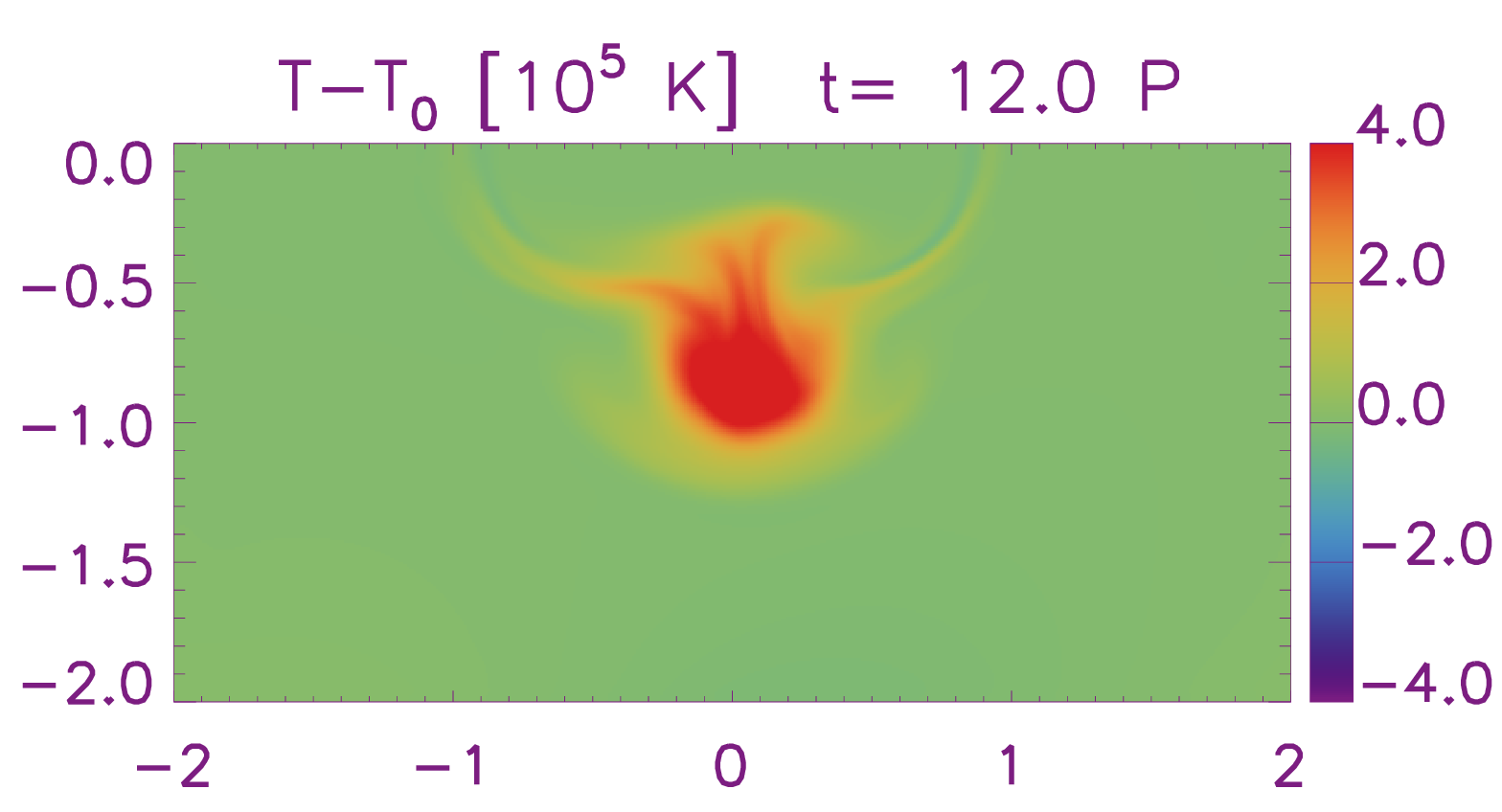} % printer
\includegraphics[scale=0.18]{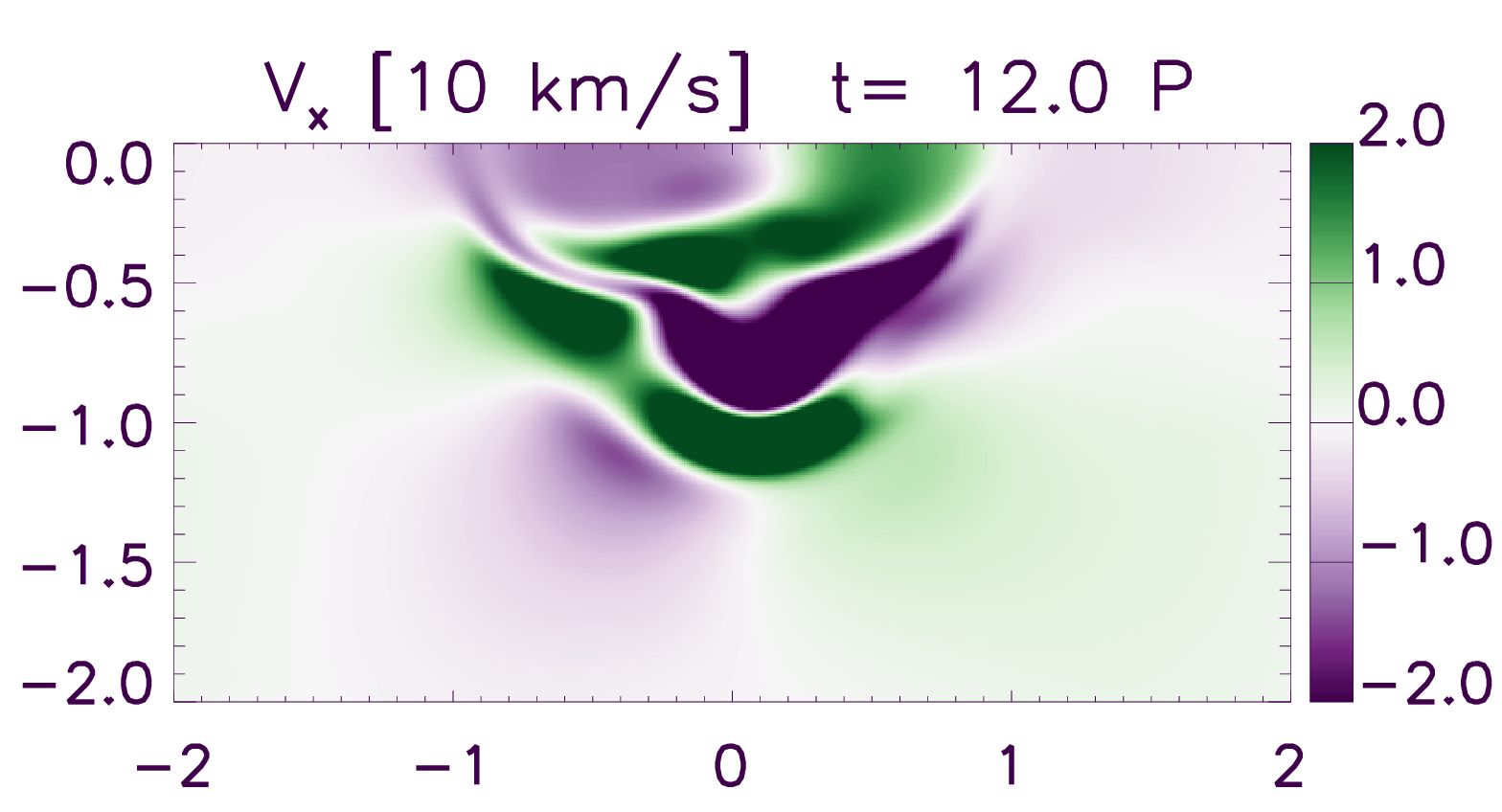} % printer
\caption{Maps on the plane $z=30$ $Mm$ of temperature difference, $\left(T(t)-T(0)\right)$ (left-hand side column),
and $v_x$ (right-hand side column) at $t=6$ $P$ (upper row) and $t=12$ $P$ (lower row)
for the simulation with continuous driver.}
\label{horiz10T}
\end{figure}
Additionally, the prolonged effect of the driver on the system
leads to the fragmentation of the structure 
of the boundary shell on the time span of a few periods.
Fig.\ref{horiz10T} shows the evolution of $v_x$
and the temperature difference $(T-T_0)$ on a cross section placed at $z=30$ $Mm$
at $t=6$ $P$ and $t=12$ $P$.
The cross sections show regular patterns at $t=6$ $P$,
when that cross section has been reached by only one pulse.
The velocity patterns are concentric around the centre and 
negative and positive velocity regions alternate.
The temperature increase is contained within an annular arc region
of the boundary shell.
Once more pulses reach this location, the 
patterns of $v_x$ and $T-T_0$ become more irregular and smaller scale structures appear.
The temperature increase extends to higher radial distances from the centre of the cylinder
and it is no longer an annular arc.
The velocity pattern also becomes irregular and involves more external shells
from the centre of the cylinder.
These structures develop when the system loses the initial symmetry about the $x=0$ plane
and analysis of the involved forces suggest that the loss of symmetry originates from the 
oscillations of the magnetic field induced by the driver.
Such an evolution is comparable with the development
Kelvin-Helmholtz instabilities as in \citet{Terradas2008}, \citet{Antolin2015}, and \citet{Magyar2016}
with the difference that in our model the conditions for development of 
the instability are built through the passage of several propagating waves,
while in those studies the instability is triggered by standing oscillations.
It has been shown that the development of Kelvin-Helmholtz instabilities
amplifies the effect of resonant absorption (or phase-mixing) of Alfv\'en waves
on plasma heating by developing smaller scale structures at the boundary shell
where wave energy is more favourably converted into heating \citep{BrowningPriest1984}.
The present simulation shows comparable dynamics,
as the thermal energy deposition in the boundary shell 
significantly accelerates after $t=9$ $P$,
when the first Kelvin-Helmholtz instability-like structures appear.
However, the high resistivity we have adopted
prohibits the development of visible small scale vortices.

\begin{figure}
\centering
\includegraphics[scale=0.4]{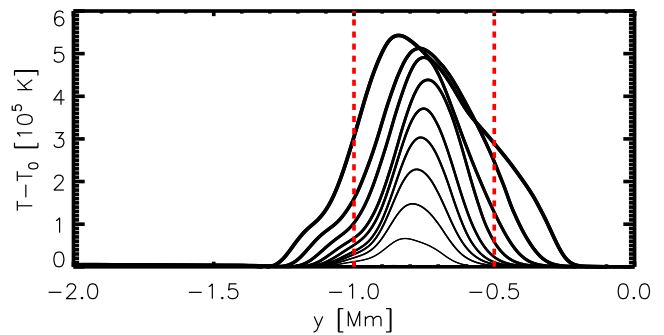}
\caption{Temperature difference, $\left(T(t)-T(0)\right)$, cut
across the cylinder on the $x=0$ plane at $z=26.60$ $Mm$
from $t=0$ to $t=12$ $P$
with $1$ $P$ cadence from the thinnest to the thickest line.
The red vertical dashed lines mark the borders of the boundary shell.}
\label{tempcut10T}
\end{figure}
One of the consequences of the expansion
of the region involved in the temperature increase is that
the heating is not bounded to the initial boundary shell, 
but it extends in time to regions initially outside of the boundary shell.
Fig.\ref{tempcut10T} shows the temperature increase on the $x=0$ plane
at $z=26.60$ $Mm$ from $t=0$ to $t=12$ $P$ with $1$ $P$ cadence 
(from the thinnest to the thickest line).
The temperature increase profile is always a curve with one peak
that reaches $0.55$ $MK$ at the end of the simulation.
We also see that the position of the centre of the peak slightly moves in time, 
but more importantly while at the beginning the heating is
only appreciable in the initial boundary shell
(between $y=[-1,-0.5]$) at the end it extends 
from $y=-1.3$ to $y=-0.2$.
This results is also in line with what has been found by \citet{Antolin2015}.

\section{Discussion and Conclusions}
\label{conclusions}
In the present work we have developed a simple model 
of a magnetised cylinder that is perturbed at one of 
its footpoints by a transverse displacement.
The configuration we have adopted leads to 
the phase mixing of propagating Alfv\'en waves
in the loop shell region
and we have focused our analysis on the heating 
that follows.
The aim of this modelling effort is to investigate
the contribution of phase-mixing of Alfv\'en waves
to the coronal heating problem.

In many respects, our model is a simplification of
a coronal loop structure and it does not address
more complex features of these magnetic structures.
Certainly, our modelling starts from a condition where 
the coronal loop is already in place and with a temperature
above the chromospheric and photospheric temperatures.
Therefore, the key to interpret our results is
whether such a mechanism can at least maintain the loop structure against
the radiative losses.

Although our study is not conclusive on the matter,
it sheds light on some relevant aspects of
the contribution from phase-mixing to coronal heating 
and it helps identify some points
where this model is in disagreement with observations.

First of all, our numerical experiments show that phase-mixing is a
plausible mechanism in the corona, where
we have used plasma and magnetic field parameters that 
are in the observed or measured range.
We have also shown that a time limited oscillation propagating
from the footpoint leads to a time limited energy deposition
that increases the plasma temperature by a certain amount.
This process is in line with the impulsive manner
how coronal heating is observed to work \citep{Warren2011,Reale2010}.
Finally, we have shown that the energy deposition is comparable
with the radiative losses thus keeping phase-mixing as a 
candidate to maintain the high temperature of the solar corona.
Given the physics and geometry of the solar corona where magnetic field
structures act as vertical wave guides
and observed horizontal motions at the base of the corona perturb these magnetic structures out of equilibrium,
inevitably leading to the upward
propagation of Alfv\'enic waves, 
phase-mixing has to occur. 
The unanswered question is whether this is just a marginal
process in the solar corona
or whether it dominates the temperature evolution.

Here, major concerns are still challenging
phase-mixing as a mechanism to justify coronal heating
on a larger scale and here we list some.
Fist of all, the model can match the radiative losses estimation
only with the adoption of a very high magnetic resistivity.
In our simulations we have used a resistivity $10^9$ times
the value predicted by the classical theory (at a 2MK temperature)
and such an amplification of the resistivity is not only essential to overcome the numerical resistivity (inherent to any MHD code)
but, more importantly, also to achieve a significant increase in the plasma temperature in the model.
Similarly, we have to adopt a relatively large amplitude driver, in order
to stimulate an appreciable deposition of thermal energy in the model.
Usually velocities of the order of 10 km/s are measured as motions of coronal structures,
while in our case the peak velocity is above 100 km/s.
Our velocity is also much higher than the one used in \citet{Pascoe2010}
and at about $10\%$ of the Alfv\'en speed,
our system probably evolves in a weakly non linear regime.
Moreover, the measured power spectrum of the oscillations of the solar corona 
peaks at 5 minutes because of the forcing photospheric oscillations.
The period of our driver is instead at 6 seconds which may not be a significant frequency of the coronal power spectrum.
This is especially relevant because longer oscillation periods would probably 
lead to slower heating time scales in this model.
Finally, our model has certainly led to the heating of the boundary shell,
but does not address how the centre of the loop would get
a significant amount of thermal energy to sustain the loop structure against 
radiative cooling.
This is a concern intrinsic to the phase-mixing process that
concentrates wave energy only on the boundary shell \citep{Cargill2016}. 
Even the single case we have addressed where the temperature increase eventually
involves regions beyond the boundary shell, it needs several 
oscillations to set in.
The question remains open
whether this extension needs to be triggered earlier 
in order to balance radiative losses, particularly in the core of the loop.

The aforementioned problems seem to pose
major concerns for the validation of phase-mixing as a mechanism
for coronal heating on a global scale.
The model we have described seems to match the observations only by pushing 
the physical parameters to conditions extremely unlikely to occur in the solar corona.
At the same time, it is useful to outline some uncertainties 
about the coronal physics that could still open the way to
the possibility that phase-mixing can explain coronal heating.
It is widely accepted that the combination of a classical resistivity value
and the spatial resolution at which we resolve coronal structure 
are not able to explain the
magnetic energy conversion in the solar corona.
One possibility is that the development of gradients on 
spatial scales much smaller than our current spatial resolution 
can affect the larger scale evolution and thus amplify the role of the classical diffusivity terms.
Another possibility is that the classical theory simply fails when small spatial scales become relevant.
In any case, the question on how the ohmic heating operates in the solar corona 
still holds and the efficiency
of any magnetic energy conversion mechanism remains uncertain.
The intensity of the driver has also been questioned, but
it is currently not possible to disentangle the line of sight projection
effects when coronal flows are measured from doppler velocities. Therefore,
while we can state that average observed speeds are of the order of $10$ $km/s$,
it is more difficult to determine whether this average
comes from a collection of comparable motions
or from the residual of the line of sight cancellation of much faster motions \citep{DeMoortelPascoe2012}.
Similarly, it needs to be addressed whether the development of Kelvin-Helmholtz instabilities
triggered by phase-mixing could make the whole mechanism more efficient.
Finally, while it is true that our model takes into account only a monochromatic
wave packet with a specific period that seems not to be relevant for the solar corona,
it is also true that such simple drivers are unlikely to occur at the dynamic base of the solar corona.
It is more plausible that actual drivers in the solar corona 
are composed of a more complex spectrum. This means that also frequencies of the order of seconds are
a component of the spectrum and that a realistic and energy richer wave packet
would provide the system with more energy 
than what we have modelled with a monochromatic pulse.

Our partial conclusion is that 
phase-mixing seems not to be a viable mechanism to
explain the large scale heating of the solar corona, even though it is likely
that some energy deposition takes place through this mechanism.
At the same time, further analysis is needed to validate the present results 
and to complete the investigation.
To begin with we will run a comparable analysis over longer time scales and with
a different class of drivers, also including in the energy balance the effects of radiative losses and
various kinds of background heating, in order to more conclusively 
address
whether phase-mixing can sustain the thermal structure of a coronal loop.

\begin{acknowledgements}
This research has received funding from the European Research Council (ERC) under the European Union's Horizon 2020 research and innovation programme (grant agreement No 647214) and from the UK Science and Technology Facilities Council.
This work used the DiRAC Data Centric system at Durham University, operated by the Institute for Computational Cosmology on behalf of the STFC DiRAC HPC Facility (www.dirac.ac.uk. This equipment was funded by a BIS National E-infrastructure capital grant ST/K00042X/1, STFC capital grant ST/K00087X/1, DiRAC Operations grant ST/K003267/1 and Durham University. DiRAC is part of the National E-Infrastructure.
We acknowledge the use of the open source (gitorious.org/amrvac) MPI-AMRVAC software, relying on coding efforts from C. Xia, O. Porth, R. Keppens.
\end{acknowledgements}

%\begin{thebibliography}{}

\bibliographystyle{aa}
\bibliography{ref}

%\end{thebibliography}

\end{document}